\documentclass[paper=a4,abstracton,numbers=noenddot]{scrartcl}
\usepackage[a4paper,left=2.5cm,right=2.5cm,top=3cm,bottom=2.5cm]{geometry}

\usepackage{authblk}

\usepackage[numbers]{natbib}

\def\tikzExtOn{}

\def\TodoNotes{}

\usepackage[intlimits]{amsmath}
\usepackage{amssymb} 
\usepackage{amsthm}
\usepackage{amsfonts,mathrsfs}
\usepackage{bm}
\usepackage{mathtools} 

\usepackage{siunitx} 

\usepackage{tikz}
\usepackage{tikz-3dplot}
\usepackage{pgfplots}
\usepackage{pgfplotstable}
\usepackage{tikzscale}
\usepackage{standalone} 

\usepackage{forloop}
\usepackage{arrayjobx}

\usepackage[noend]{algorithmic}
\usepackage{algorithm}

\usepackage{booktabs} 
\usepackage{hyperref} 
\usepackage{xspace} 

\usepackage{xcolor,pdfcomment,soul} 
\usepackage[textsize=footnotesize]{todonotes}

\usepackage{caption}
\usepackage{subcaption} 			
\usepackage[labelformat=simple]{subcaption}   	

\usepackage{enumerate} 
\usepackage{enumitem}
\usepackage{cleveref}
\crefformat{figure}{Figure~#2#1#3} 
\Crefformat{figure}{Figure~#2#1#3} 
\crefformat{equation}{(#2#1#3)} 
\Crefformat{equation}{(#2#1#3)} 
\labelcrefformat{equation}{(#2#1#3)}  
\crefformat{section}{Section~#2#1#3} 
\Crefformat{section}{Section~#2#1#3} 
\crefformat{appendix}{#2#1#3} 

\usepackage[utf8]{inputenc} 
\usepackage[german,english]{babel}  
\usepackage[T1]{fontenc} 
\usepackage{lmodern} 

\newcommand{\bmchange}[1]{#1} 
\newcommand{\rhchange}[1]{#1} 
 
\def\Black{\color{black}}
\def\Blue{\color{black}}

\newcommand{\myMat}[1]{\mathbf{#1}}

\newcommand\indexA{i} 		
\newcommand\indexB{j} 		
\newcommand\indexC{k} 		

\newcommand\indexSpan{s}	

\newcommand{\Bspline}{B} 		
\newcommand{\KV}{\varXi} 		
\newcommand{\KVRefined}{\tilde{\varXi}} 	
\newcommand\uu{\xi} 			
\newcommand\uuRefined{\tilde{\uu}} 	
\newcommand\UVsurf{\pt{\uu}} 	
\newcommand\multi{m} 			

\newcommand{\CPara}{c}			

\newcommand{\BsplineSpace}{\mathbb{S}} 

\newcommand\pu{p}			
\newcommand\pdim{n_{sd}}			

\newcommand{\iSubDim}{d}

\newcommand{\supportdomain}{\mathcal{S}^{\textnormal{v}}}
\newcommand{\supportdomainReg}{\mathcal{S}^{\textnormal{r}}}
\newcommand{\supportdomainCut}{\mathcal{S}^{\textnormal{c}}}
\newcommand{\supportdomainDWQ}{\mathcal{S}^{\textnormal{WQ}}}
\newcommand{\supportdomainGQ}{\mathcal{S}^{\textnormal{GQ}}}
\newcommand{\supportdomainDWQglobal}{\mathcal{S}^{\patchdomain_{\textnormal{WQ}}}}

\newcommand{\visibledomain}{\patchdomain} 
\newcommand{\TrimCurve}{\Gamma}     

\newcommand{\domain}{\Omega}
\newcommand{\patchdomain}{\Omega^{\textnormal{v}}}
\ifcsname boundary\endcsname%
  
\else%
  
\fi%

\newcommand{\R}{\mathbb{R}}

\DeclareMathOperator{\supp}{supp}

\DeclareMathOperator{\diag}{diag}

\newcommand{\breakpoint}{{\uu}^{b}}
\newcommand{\trialFct}{B}
\newcommand{\testFct}{M}
\newcommand{\targetFct}{\trialFct^{*}}
\newcommand{\quadWeight}{w}
\newcommand{\quadPoint}{x}
\newcommand{\quadRule}{\mathbb{Q}}
\newcommand{\quadIndexSet}{\mathcal{Q}} 
\newcommand{\discCoeff}{{\uu}^{\textnormal{disc}}} 
\newcommand{\quadRuleDisc}{\mathbb{DQ}}

\newcommand{\OfOrder}{\mathcal{O}}
\providecommand\url[1]{\emph{#1}}

\newcommand\XXint[3]{{\setbox0=\hbox{$#1{#2#3}{\int}$}
    \vcenter{\hbox{$#2#3$}}\kern-.5\wd0}}

\newcommand\fromto[2]{\{ #1, \dots, #2 \}}

\newcommand\pt[1]{\boldsymbol{#1}}

\ifdefined\tikzSubmission

\else

\fi

\ifdefined\tikzSubmission

\else

\fi

{
  \begin{table}[#1]
    \def\mycap{#2}
    \def\mylabel{#3}
    \centering
    \begin{tabular}{#4}
      \toprule
}
{
  \bottomrule
  \end{tabular}
  \caption{\mycap}
  \label{\mylabel}
  \end{table}
}

{
  \begin{table}[#4]
    \def\mycap{#1}
    \def\mylabel{tab:#2}
    \centering
    \begin{tabular}{#3}
      \toprule
}
{
  \bottomrule
  \end{tabular}
  \caption{\mycap}
  \label{\mylabel}
  \end{table}
}

\ifdefined\TodoNotes
    \newcommand{\mycomment}[2][-] 
    {
        {%
            \ifdefined\tikzExtOn
            \tikzset{external/export=false}
            \fi
            \todo[color={green!33}]{ \textbf{[\uppercase{#1}]} #2}
            \pdfcomment[color=yellow,opacity=0.0,author=#1]{}
        }%
    }	
    \newcommand{\myhighlight}[2][-] 
    {
    \pdfcomment[color=yellow,voffset=-5pt,hoffset=3pt,opacity=0.0,author=#1]{}
    \hl{#2}%
    }	
    \setstcolor{red} 
    \newcommand{\mydelete}[2][-] 
    {
    \pdfcomment[color=yellow,voffset=-5pt,hoffset=3pt,opacity=0.0,author=#1]{}
    \st{#2}%
    }
\else
    \newcommand{\mycomment}[2][-]{}
    \newcommand{\myhighlight}[2][-]{}
    \newcommand{\mydelete}[2][-]{}
\fi

\ifdefined\tikzSubmission
\newcommand\revdel[2]{}
\newcommand\revadd[2]{}
\newcommand\revmod[2]{}
\newcommand\revcomment[1]{} 
\else
\newcommand\revdel[2]{ %
{
  \pdfmarkupcomment[markup=StrikeOut,color=red,author=Marussig and Zechner]{#1}{#2}
}
}
\newcommand\revadd[2]{ %
{
  \pdfmarkupcomment[markup=Underline,color=green,author=Marussig and Zechner]{#1}{#2}
}
}
\newcommand\revmod[2]{ %
{
  \pdfmarkupcomment[markup=Highlight,color=yellow,author=Marussig and Zechner]{#1}{#2}
}
}
\newcommand\revcomment[1]{ %
{
  \pdfcomment[color=yellow,author=Marussig and Zechner,voffset=8pt,opacity=0.8]{#1}
}
}
\fi

\newtheoremstyle{myremark}
{3pt}
{3pt}
{}
{}
{\itshape}
{:}
{.5em}
{}
\theoremstyle{myremark}
\newtheorem*{remark}{Remark}

\newcounter{footnoteNumber} 

\usepackage{xfrac}

\tikzset{%
  highlight/.style={rectangle,rounded corners,fill=red!60,draw,fill opacity=0.125,thick,inner sep=0pt}
}

\usepackage{colortbl}

\newcommand{\vect}[1]{\boldsymbol{#1}} 	
\newcommand{\idx}[1]{\boldsymbol{#1}} 

\newcommand\ndofs{n}
\newcommand\ndofsRefine{\tilde{n}}

\newcommand\x{\uu}
\newcommand\p{\pu}

\newcommand{\Supp}[1]{\ensuremath{\mathrm{supp}\left\{#1\right\}}}

\newcommand\subDMatrix{\myMat{S}}

\title{Fast immersed boundary method based on weighted quadrature}

\begin{document}

\author[1]{Benjamin Marussig}%
\author[2]{Ren\'{e} Hiemstra}
\author[2]{Dominik Schillinger}
\affil[1]{Institute of Applied Mechanics, Graz University of Technology}
\affil[2]{Institute for Mechanics, Technical University of Darmstadt}
\date{}
\maketitle                   

\begin{abstract}
    \bmchange{Combining sum factorization, weighted quadrature, and row-based assembly enables} efficient higher-order computations for tensor product splines.
    We aim to transfer these concepts to immersed boundary methods, which perform simulations on a regular background mesh cut by a boundary representation that defines the domain of interest. 
    Therefore, we present a novel concept to divide the support of cut basis functions to obtain regular parts suited for sum factorization. These regions require special discontinuous weighted quadrature rules, while Gauss-like quadrature rules integrate the remaining support.
    Two linear elasticity benchmark problems confirm the derived estimate for the computational costs of the different integration routines and their combination.
    Although the presence of cut elements \bmchange{reduces} the speed-up, its contribution to the overall computation time declines with \bmchange{$h$-refinement}. 
\end{abstract}




\section{Introduction}
In the field of finite element analysis, the generation of a boundary-conforming mesh presents challenges, particularly for complex 3D geometries. This process that often requires labor-intensive manual intervention hinders the efficiency of the design-to-analysis workflow, driving current research and development towards more efficient interactions between geometric models and finite element analysis. Over the past two decades, two families of methods have emerged to address this challenge: isogeometric analysis and immersed boundary methods.

Isogeometric analysis (IGA), introduced in 2005 \cite{Hughes2005}, aims to bridge the gap between computer-aided design and finite element analysis. The initial techniques \cite{Hughes2005,Cottrell2006,Cottrell2007,Cottrell2009} combined spline technology from the field of computer-aided design with standard finite element formation, assembly, and solution procedures. It was soon recognized that the smoothness and structure of splines enable more efficient implementations than were previously possible in IGA and standard $C^0$-continuous finite element analysis alike \cite{hughes2010efficient,Schillinger2014,gao2014fast,mika2021matrix}. Techniques such as sum factorization, a classical approach in \textit{hp}-finite element methods that leverages the local tensor product structure, have been employed to expedite the element formation process \cite{Antolin2015}. Additionally, more efficient quadrature techniques have been developed, including Generalized Gaussian quadratures \cite{Hiemstra2017,Johannessen2017}, reduced integration \cite{Schillinger2014,Hiemstra2017}, and weighted quadrature \cite{Calabro2017,sangalli2018matrix,Hiemstra2019,Giannelli2022a}. Studies such as \cite{Calabro2017,Hiemstra2019} have demonstrated that a combination of sum factorization, weighted quadrature, and row-based assembly can lead to significant speed-ups in the formation and assembly of system matrices. Notably, these advancements have enabled efficient higher-order computations, with the number of operations scaling as $\mathcal{O}(p^4)$ instead of $\mathcal{O}(p^9)$, as observed in classical techniques \bmchange{\cite{Calabro2017,Hiemstra2019}}. Despite the success of IGA as an analysis technology, it has not fully resolved the challenges associated with mesh generation, leaving a significant gap between computer-aided design and finite element analysis unresolved.

Immersed boundary methods, also known as fictitious domain, embedded domain or cut finite element methods, eliminate the need for boundary-conforming discretizations altogether. Thus, they mitigate the complexities of meshing procedures and frequent grid regeneration required for models involving substantial deformations and displacements. Immersed boundary methods offer an alternative to boundary-conforming meshes, but introduce other computational challenges: (1) numerical evaluation of integrals over cut elements, (2) imposition of boundary conditions on immersed boundaries, and (3) maintaining stability of discrete function spaces, in particular in the presence of very small cut elements. 
Over the past decades, numerous advanced techniques have been developed to address these challenges. As a result, various variants of immersed boundary methods exist, including the finite cell method \cite{parvizian2007finite}, certain versions of the extended finite element method \cite{gerstenberger2008extended,haslinger2009new}, isogeometric immersed boundary methods \cite{schillinger2012isogeometric}, fixed-grid methods in fluid-structure interaction \cite{kamensky2015immersogeometric,xu2018framework}, \bmchange{weighted extended B-splines \cite{Hoellig2003,Hoellig2003a,Chu2022a}}, and embedded domain approaches utilizing penalty methods \cite{breitenberger2015analysis}, Lagrange multipliers \cite{glowinski1999distributed,burman2010fictitious,kamensky2017projection}, Nitsche-type methods \cite{hansbo2002unfitted,embar2010imposing,ruess2014weak}, and discontinuous Galerkin methods \cite{schillinger2016non,gurkan2019stabilized}. 
It is important to note that this list is not exhaustive, and numerous other variations of immersed boundary methods are available. For a comprehensive overview, we refer the interested reader to the reviews articles \cite{schillinger2015finite,burman2015cutfem,de2023stability} and the references therein. 
\bmchange{It is also worth noting that there is a close relation to the treatment of trimmed domains as they occur in CAD geometries \cite{Marussig2018}.}

In this paper, we draw our attention to another challenge of the isogeometric immersed boundary method that uses higher-order smooth spline basis functions: applying efficient implementation concepts. Transferring efficient implementation concepts that work for boundary-conforming tensor product spline discretizations is usually not possible, because the arbitrarily located immersed boundary destroys the smoothness and structure of the background mesh's spline discretization. To overcome this obstacle, a few approaches have been suggested in the literature. In \cite{Messmer2022}, a partitioning into macro-elements is proposed. 
Since \bmchange{these} elements follow the tensor product structure,
Generalized Gaussian quadrature can be used within these subregions. 
\bmchange{Alternatively, such rules may be employed for all non-cut basis functions, while cut ones are integrated with Gauss quadrature, as recently suggested in \cite{loibl2023patchwise}. Due to the overlap of the supports of cut and non-cut basis functions, this procedure leads to transition elements that require reduced quadrature and Gauss points.}
Another approach proposed by the first author allows the utilization of weighted quadrature, sum factorization, and row assembly
-- a combination we will refer to as fast formation and assembly --
by introducing so-called discontinuous weighted quadrature rules \cite{Marussig2022,Marussig2022a}. 
Up to now, this concept has only been tested for simple $L^2$-projection problems.
Therefore, in this paper, we extend this concept to linear elasticity problems by combining it with the ideas presented in \bmchange{\cite{Hiemstra2019}}.
Furthermore, we propose a novel strategy to set up more efficient discontinuous weighted quadrature rules and estimate the associated computational costs in terms of the number of floating-point operations.

\bmchange{
\section{Preliminaries}
}
In this work, a \bmchange{tensor product} spline discretization defines the background mesh.
For a detailed discussion on splines, we refer to \cite{Cohen2001,Cottrell2009,Boor2001} and restrict ourselves here to a few preliminaries required later on.
A spline is a set of piecewise polynomial segments with prescribed continuity at their breakpoints.  
The corresponding basis functions are \emph{B-splines} $\hat{\Bspline}_{\indexA,\pu}$, 
which specify the spline by its degree $\pu$ and a non-decreasing sequence $\KV$ of parametric coordinates~{$\hat{\uu}_\indexB \leqslant \hat{\uu}_{\indexB+1}$} called knot vector and knots, respectively.
Each knot value marks a breakpoint $\breakpoint_\indexC$ of the spline, and the related knot-multiplicity $\multi\left(\breakpoint_\indexC\right)$ specifies the continuity $C^{r_\indexC}$ at $\breakpoint_\indexC$ by $r_\indexC={\pu-m\left(\breakpoint_\indexC\right)}$. 
To obtain splines with maximal smoothness, $\multi=1$ for all interior knot values.
It is often convenient to have $C^{-1}$ continuity at the splines boundary.
This property is accomplished by using so-called open knot vectors, which are characterized by setting the multiplicity of end knot values to $\multi=p+1$. 
In general, a knot vector $\KV$ defines an entire set of linearly independent B-splines~\bmchange{$\{\hat{\Bspline}_{\indexA,\pu}\}_{\indexA=1}^{\ndofs}$} on the parametric domain $\hat{\domain}$.
The corresponding space of splines on an interval $[a,b]$ is given by
\begin{align}
    \label{eq:splineSpace}
    \BsplineSpace^{\pu}_{\boldsymbol{r}}
    \left( [a,b] \right) := \left\{ \left. \sum_{\indexA}  \hat{\Bspline}_{\indexA,\p}(\x) \CPara_\indexA \;  \right| \;  \x \in [a,b], \; \CPara_\indexA \in \mathbb{R}, \quad \indexA=\bmchange{1,\dots,\ndofs} \right\}.
\end{align}
where $\boldsymbol{r}$ is a collection of all regularities $r_\indexC$ at the breakpoints $\breakpoint_\indexC$. 
Each $\hat{\Bspline}_{\indexA,\pu}$ has local support, $\supp{ \{\hat{\Bspline}_{\indexA,\pu} \} }$, specified by the knots $\fromto{\hat{\uu}_{\indexA}}{\hat{\uu}_{\indexA+\pu+1}}$. 
Multivariate basis functions $\hat{\Bspline}_{\boldsymbol{\indexA},\boldsymbol{\pu}}$ of dimension $\bmchange{\pdim}$ are obtained by computing the tensor product of univariate B-splines $\hat{\Bspline}_{\indexA_{\bmchange{\iSubDim}},\pu_{\bmchange{\iSubDim}}}$ defined by separate degrees $\pu_{\bmchange{\iSubDim}}$ and knot vectors $\KV_{\bmchange{\iSubDim}}$ for each parametric direction $\bmchange{\iSubDim}=1,...,\bmchange{\pdim}$.
Thus the evaluation of a multivariate B-spline at a parametric coordinate $\UVsurf=(\uu_1,\dots,\uu_{\bmchange{\pdim}})$ can be generally expressed as
\begin{align}
    \label{eq:tensorProductBspline}
    \hat{\Bspline}_{\boldsymbol{\indexA},\boldsymbol{\pu}}(\UVsurf)=\bmchange{\prod^{{\pdim}}_{{\iSubDim}=1} \hat{\Bspline}_{\indexA_{{{\iSubDim}}},\pu_{{\iSubDim}}}(\uu_{{\iSubDim}})}.
\end{align}

A property that will be useful later is that the first derivative of a B-spline $\hat{\Bspline}^{(1)}_{i,p}$ can be expressed by a linear combination of B-splines of the previous degree
\begin{align}
	\label{eq:Bspline_Np_Deriv}
	\hat{\Bspline}^{(1)}_{i,p}(\uu)  
    = \frac{p}{\hat{\uu}_{i+p}-\hat{\uu}_{i}} \: \hat{\Bspline}^{(0)}_{i,p-1}(\uu) 
	- \frac{p}{\hat{\uu}_{i+p+1}-\hat{\uu}_{i+1}} \: \hat{\Bspline}^{(0)}_{i+1,p-1}(\uu), && \textnormal{where}&& \hat{\Bspline}_{j,p-1}^{(0)}\coloneqq\hat{\Bspline}_{j,p-1}.
\end{align}

Usually, $\hat{\Bspline}_{i,p}$ denotes a B-spline defined in the parameter space $\hat{\domain}$, while ${\Bspline}_{i,p}$ represents its counterpart mapped to the physical space $\domain$.
In the case of immersed boundary methods, the geometric mapping is often the identity.
Hence, the background mesh and the parameter space coincide, allowing us to skip the hat-notation for the remainder of the paper.

\section{Weighted quadrature}

We \bmchange{shall use} \emph{weighted quadrature} (WQ) for the formation of mass and stiffness matrices. 
Hence, we consider the following univariate \bmchange{integrals}
\begin{align}
    \label{eq:integralOfInterest}
    &\int_\domain \testFct^{(0)}_\indexA(\uu) \trialFct^{(0)}_\indexB(\uu) \CPara(\uu) d\uu, && &&\int_\domain \testFct^{(0)}_\indexA(\uu)        \trialFct^{(1)}_\indexB(\uu) \CPara(\uu) d\uu, \\
    \label{eq:integralOfInterestDeriv}
    &\int_\domain \testFct^{(1)}_\indexA(\uu) \trialFct^{(0)}_\indexB(\uu) \CPara(\uu) d\uu, && &&
    \int_\domain \testFct^{(1)}_\indexA(\uu) \trialFct^{(1)}_\indexB(\uu) \CPara(\uu) d\uu. 
\end{align}
$\testFct_\indexA(\uu)$ and $\trialFct_\indexB(\uu)$ are test and trial functions in the spline space $\BsplineSpace^{\pu}_{\boldsymbol{r}}$, and $\CPara(\uu)$ is determined by the geometry mapping and the material behavior.
The integrals above can be concisely written as 
\begin{align}
    \label{eq:integralOfInterestShort}
    \int_\domain \testFct^{(\alpha)}_\indexA(\uu) \trialFct^{(\beta)}_\indexB(\uu) \CPara(\uu) d\uu && \textnormal{with} && \alpha, \beta = 0,1.
\end{align}
In general, numerical quadrature rules are designed to be exact for the case that $\CPara(\uu)=1$.
Considering weighted quadrature, quadrature rules $\quadRule^{(\alpha)}_\indexA$ are designed for each test function and its derivative $\testFct^{(\alpha)}_\indexA$ by incorporating them into the quadrature weights
\begin{align}
    \label{eq:weightedQuadRule}
    \quadRule^{(\alpha)}_\indexA = \sum_\indexC \trialFct^{(\beta)}_\indexB(\quadPoint_\indexC)  \quadWeight^{(\alpha)}_{\indexC,\indexA} \coloneqq \int_\domain \trialFct^{(\beta)}_\indexB(\uu)  \left(\testFct^{(\alpha)}_\indexA(\uu)  d\uu \right) && \textnormal{with} && \alpha, \beta = 0,1.
\end{align}
Given a suitable layout of quadrature points $\quadPoint_\indexC$, 
which will be discussed later on in \cref{sec:layout}, the weights $\quadWeight^{(\alpha)}_{\indexC,\indexA}$ can be computed by solving the following system of equations
\begin{align}
    \begin{aligned}
        \label{eq:weightedQuadRuleSystem}
        \quadRule^{(\alpha)}_\indexA\left(\targetFct_{\indexB_1}\right) &=&& \sum_{\indexC\in\quadIndexSet_\indexA} \targetFct_{\indexB_1}(\quadPoint_\indexC)  \quadWeight^{(\alpha)}_{\indexC,\indexA}&&\coloneqq \int_\domain \targetFct_{\indexB_1}(\uu)  \left(\testFct^{(\alpha)}_\indexA(\uu)  d\uu \right) \\
        &\vdotswithin{=}&&  &&\vdotswithin{:=}\\
        \quadRule^{(\alpha)}_\indexA\left(\targetFct_{\indexB_{\ell}}\right) &=&& \sum_{\indexC\in\quadIndexSet_\indexA}  \targetFct_{\indexB_{\ell}}(\quadPoint_\indexC)  \quadWeight^{(\alpha)}_{\indexC,\indexA}&&\coloneqq \int_\domain \targetFct_{\indexB_{\ell}}(\uu)  \left(\testFct^{(\alpha)}_\indexA(\uu)  d\uu \right) 
    \end{aligned}
\end{align}
where \bmchange{$\targetFct_{\indexB}$} are the B-splines of the target space $\BsplineSpace^{\pu}_{\boldsymbol{r}-1}$ \bmchange{whose support intersects that of}  
the test \bmchange{function $\testFct^{(\alpha)}_{\indexA}\in \BsplineSpace^{\pu}_{\boldsymbol{r}}$}.
Note that the spline spaces of the trial functions and their derivatives are contained within the target space since $\BsplineSpace^{\pu}_{\boldsymbol{r}}\subset\BsplineSpace^{\pu}_{\boldsymbol{r}-1}$ and $\BsplineSpace^{\pu-1}_{\boldsymbol{r}-1}\subset\BsplineSpace^{\pu}_{\boldsymbol{r}-1}$.
The indices $\indexB_1,\dots,\indexB_{\ell}$ refer to all target functions whose support overlaps with the one of the current test function, $\supp{ \{\testFct^{(\alpha)}_\indexA \} }$, and the index set $\quadIndexSet_\indexA$ refers to all quadrature points that lie within $\supp{ \{\testFct^{(\alpha)}_{\indexA} \} }$.
The system \labelcref{eq:weightedQuadRuleSystem} can be solved for  $\quadWeight^{(\alpha)}_{\indexC,\indexA}, \forall k\in \quadIndexSet_\indexA$ by QR-factorization but the solution maybe not unique \cite{Hiemstra2019}.
\bmchange{Using QR-factorization is equivalent to solving the least norm solution in the discrete $l^2$-norm.
    Positivity of the weights is not explicitly enforced in the optimization problem. Therefore, negative weights can occur. In practice, this is not a large issue because the quadrature rules attain full accuracy, that is, all splines in the target space are integrated exactly to machine precision.}
In the case of multivariate tensor product test functions, quadrature rules $\quadRule^{(\alpha_1)}_{\indexA_1},\dots,\quadRule^{(\alpha_{\bmchange{\pdim}})}_{\indexA_{\bmchange{\pdim}}}$ are computed for each parametric direction.

\bmchange{
In the remainder of the paper, the test functions $\testFct_\indexA$ and trial functions $\trialFct_\indexA$ are the same. Thus, we will generally refer to both function types as $\trialFct_\indexA$ from now on.
}

\subsection{Layout of quadrature points}
\label{sec:layout}

We consider weighted quadrature rules with predefined point locations $x_k$, which allows the computation of the associated quadrature weights by solving \cref{eq:weightedQuadRuleSystem}. 
The \bmchange{total} number of weighted quadrature points $n_q$ and their distribution must be set such that this
system of equations \cref{eq:weightedQuadRuleSystem} is well defined \bmchange{for all test functions}. 
That is, the resulting system matrix is of full rank, and $n_q$ is equal to or greater than the number of exactness conditions enforced.

In general, the required $n_q$ increases when the smoothness of the spline space reduces. 
In the extreme case where the continuity reduces to $C^0$ everywhere, the minimum number of weighted quadrature points equals the number of Gauss points.
For spline spaces with arbitrary continuity, a general procedure is presented in \cite{Hiemstra2019} to determine $n_q$ and a suitable point distribution.
For \bmchange{splines} with maximal smoothness, $p+1$ points are required in elements next to discontinuities, e.g., at the boundary of open knot vectors.
For the remaining inner elements, on the other hand, the number of required quadrature points $n^i_q$ is $2$ when setting up a mass matrix and $3$ when setting up a stiffness matrix. 

\begin{figure}[ht]
    \centering
    \includegraphics[scale=1.0]{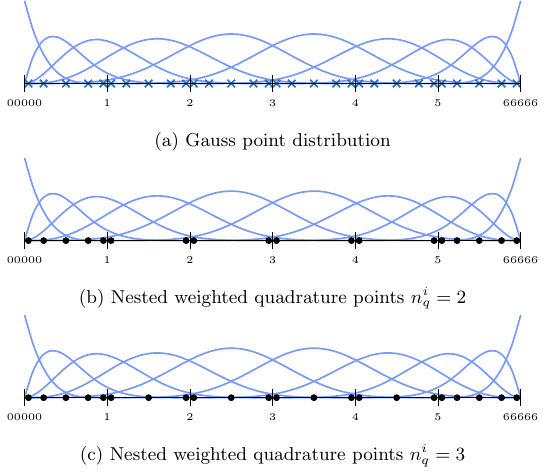}
    \caption{Weighted quadrature layout for a spline space with $p=4$ and maximal smoothness: (a) required Gauss points (blue crosses) within each element serve as a superset for (b) the weighted quadrature points (black circles) \bmchange{for setting mass matrices and (c) stiffness matrices.}}
    \label{fig:DWQlayout}
\end{figure}
The location of the weighted quadrature points within an element is arbitrary as long as \bmchange{they} do not coincide.
Often a uniform distribution is chosen, which may or may not contain the boundary of the element. 
In this work, we do not include boundary points because the proper assembly of their integral contributions increases the implementation complexity when there is more than one quadrature rule per dimension.
\bmchange{Here, the critical part is to correctly assign the contribution of the boundary points to the test functions that employ different integration rules and, therefore, may share only some of these points.} 
We will pick the weighted quadrature points as a subset of the locations of the standard element-wise Gauss quadrature points, as illustrated in \cref{fig:DWQlayout}.
\bmchange{If $p$ is the polynomial degree of the test and trial spaces, we require Gauss rules with $\pu+1$ points, which is also the maximal number of weighted quadrature points needed per element. For the interior elements of maximally smooth splines, we pick the outer Gauss points and add the one in the middle if an odd number of weighed quadrature points is needed. This choice is random, but note that the points for setting up a mass matrix (\cref{fig:DWQlayout}(b)) and a stiffness matrix (\cref{fig:DWQlayout}(c)) coincide, which allows us to reuse point evaluations if the simulation requires mass and stiffness matrices.}
\Cref{sec:detectdisc} will clarify the motivation for \bmchange{using the Gauss point layout}; in principle, it \bmchange{again} allows the reuse of evaluations at the quadrature points\bmchange{, but this time for the case} when the numerical integration performed over an element employs weighed and Gauss rules.

\section{Discontinuous weighted quadrature} \label{sec:DWQ}

\bmchange{This section details the extension of weighted quadrature to the immersed boundary method. In particular, the different types of basis functions within cut background meshes and their integration are discussed. Then, we focus on deriving weighted quadrature rules for test functions cut by the boundary $\TrimCurve$ and estimate the related computational cost.}

\subsection{Function types of cut background meshes}
\label{sec:functionTypes}

\begin{figure}[t]
    \centering
    \includegraphics[scale=0.8]{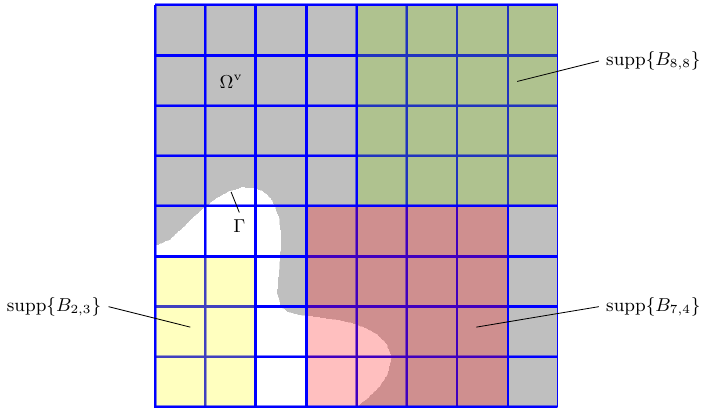}
    \caption{Background mesh defined by a bi-cubic B-spline basis with the boundary $\TrimCurve$ specifying the valid domain $\patchdomain$ (gray). Examples for the resulting \bmchange{B-spline types} based on the overlap of the support, $\Supp{ \Bspline_{\indexA_1,\indexA_2} }$, with $\patchdomain$: interior (green), cut (red), and exterior (yellow).}
    \label{fig:basisFunctionTypes}
\end{figure}

The boundary $\TrimCurve$ splits the background mesh into the exterior region and the domain of interest $\patchdomain$ for the computation.
Thus, $\patchdomain$ consists of \emph{cut elements} and regular \emph{interior elements}.
Likewise, the support of a B-spline  $\Bspline_{{\boldsymbol{\indexA}}}$  within the background mesh may also be restricted based on its overlaps with $\patchdomain$, i.e., $\supportdomain_{\boldsymbol{\indexA}} \coloneqq  \Supp{ \Bspline_{{\boldsymbol{\indexA}}} }  \cap \Blue {\patchdomain} \Black$.
This measure leads to the classification of three different basis function types:
\begin{itemize}
    \item \emph{Exterior} if $\supportdomain_{\boldsymbol{\indexA}} = \emptyset$, 
    \item \emph{Interior} if $\supportdomain_{\boldsymbol{\indexA}} = \Supp{ \Bspline_{{\boldsymbol{\indexA}}} }$, 
    \item \emph{Cut} if  {$0 < \left|\supportdomain_{\boldsymbol{\indexA}}\right| <\left| \Supp{ \Bspline_{{\boldsymbol{\indexA}}} }\right|$,} 
\end{itemize}
where $\left|\cdot\right|$ denotes the Lebesgue measure in $\R^{\bmchange{\pdim}}$.
\cref{fig:basisFunctionTypes} shows examples of these different types.
%
Exterior B-splines do not contribute to the system of equations, 
and interior ones can be treated directly by standard procedures, which is, in our case, the fast formation and assembly with weighted quadrature. Cut basis functions, however, cannot utilize these weighted quadrature rules directly since they do not consider the 
boundary $\TrimCurve$. In the following, we discuss the efficient numerical integration of the cut basis functions and the construction of suitable weighted quadrature rules.

\subsection{Integration of cut basis functions}

We aim at applying weighted quadrature to cut basis functions. 
Thereby, the main difficulty is that the boundary $\TrimCurve$ introduces an arbitrarily located jump discontinuity within the background mesh.
Integrating without taking this interface into account or simply neglecting quadrature points outside of the valid domain $\visibledomain$ will lead to incorrect results. 
Consequently, the quadrature rule has to account for these arbitrarily located discontinuities.
Considering weighted quadrature, this circumstance affects the correct representation of the integration domain $\supportdomain_{\boldsymbol{\indexA}}$ of a cut B-spline $\Bspline_{\boldsymbol{\indexA}}$ and the computation of its weighted quadrature rules. 
\bmchange{Besides}, sum factorization cannot be applied \bmchange{since} $\supportdomain_{\boldsymbol{\indexA}}$ does not follow a tensor product structure in general. 

In a first step, we split the domain $\supportdomain_{\boldsymbol{\indexA}}$ into a \emph{regular} part $\supportdomainReg_{\boldsymbol{\indexA}}$, which follows the tensor product structure (at least on the element-level), and a \emph{cut} part $\supportdomainCut_{\boldsymbol{\indexA}}$, which consists of all elements cut by the boundary. Hence, the integral over a cut basis function can be written as
\begin{align}
    \label{eq:domainSplitting}
    \int_{\supportdomain_{\boldsymbol{\indexA}}}    \Bspline_{\boldsymbol{\indexA}}(\UVsurf) d\UVsurf = 
    \int_{\supportdomainReg_{\boldsymbol{\indexA}}} \Bspline_{\boldsymbol{\indexA}}(\UVsurf) d\UVsurf + 
    \int_{\supportdomainCut_{\boldsymbol{\indexA}}} \Bspline_{\boldsymbol{\indexA}}(\UVsurf) d\UVsurf. 
\end{align}

The integration of $\supportdomainCut_{\boldsymbol{\indexA}}$ requires an element-wise quadrature rule that can treat arbitrary interfaces within an element.
There is indeed a large body of literature proposing approaches for this task.
One group of schemes represents cut elements by sub-elements that allow the utilization of conventional -- usually Gaussian -- quadrature rules, e.g., \cite{Antolin2019,cheng2010,Fries2016,Kudela2013,Kudela2016,Kudela2015,legay2005}.
Such a \bmchange{reparameterization} may also involve a mapping of the background mesh \cite{Lehrenfeld2016,Lehrenfeld2021}.
Another strategy is the construction of tailored integration rules for each cut element, e.g., \cite{mousavi2011,Mueller2013,Nagy2015,Saye2015,Saye2022,Gunderman2021,Gunderman2021a}.
These strategies usually obtain integration weights by solving a system of moment-fitting equations.
When choosing the approach best suited for a specific application, the first distinguishing feature is if the interface $\TrimCurve$ is defined in an implicit or a parametric representation. 
In this work, a level set function specifies $\TrimCurve$, and we employ the algorithms for implicitly defined geometry (Algoim) presented in \cite{Saye2015,Saye2022} for the integration of the cut elements representing $\supportdomainCut_{\boldsymbol{\indexA}}$.

\begin{figure}[ht]
    \centering
    \includegraphics[scale=1.0]{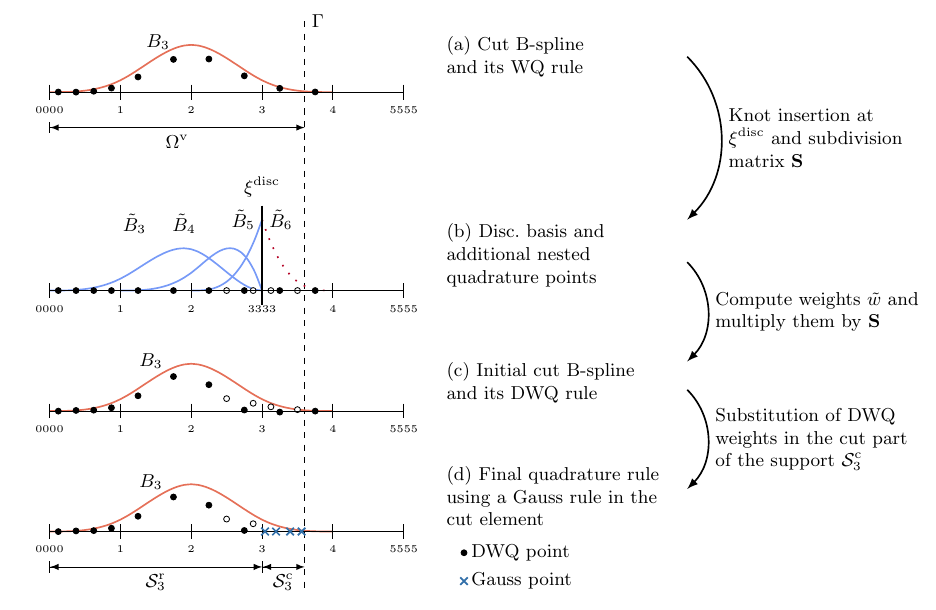}
    \caption{Construction steps of a DWQ rule for integrating a mass matrix of a cubic basis cut at position $\TrimCurve$ shown for the B-spline $\Bspline_{3,3}$: (a) Conventional weighted quadrature (WQ) points (black dots) for $\Bspline_{3,3}$. (b) Quadrature layout with additional quadrature points (white) for WQ of the refined discontinuous $\tilde{\Bspline}_{\indexB,3}$ associated with $\Bspline_{3,3}$. (c) Linear combination of the refined discontinuous WQ rules to obtain the DWQ for $\Bspline_{3,3}$. In (a,c), the points' height indicates the related weight value. Knot insertion allows going from (a) to (b), and applying the related subdivision matrix $\mathbf{S}$ allows the mapping from (b) to (c). }
    \label{fig:DWQconstr}
\end{figure}

\begin{remark} 
    In the subsequent \cref{fig:DWQconstr,fig:DWQlocal,fig:DWQglobal,fig:DWQglobalExamples}, the quadrature points in cut elements are defined by Gauss rules mapped by sub-elements and not by the integration rules \cite{Saye2015,Saye2022} utilized in the numerical experiments.
\end{remark}

\bmchange{
\subsection{Computing univariate discontinuous quadrature rules}

This section details the computation of weighted quadrature rules applicable to cut test functions. 
Here, the focus lies on the univariate setting 
to present the essential procedure for obtaining suitable quadrature points and weights. Later on, \cref{sec:detectdisc} discusses the required extension to the multivariate case.

\subsubsection{Computation of discontinuous weighted quadrature rules $\quadRuleDisc^{(0)}(\cdot)$}
}
\label{sec:DWQcomputation}

The domain splitting\bmchange{, $\supportdomain_{{\indexA}}=\supportdomainReg_{{\indexA}}\cup\supportdomainCut_{{\indexA}}$,} allows us to extract the regular integration region \bmchange{$\supportdomainReg_{{\indexA}}$, in which the corresponding elements are not cut.}
However, standard weighted quadrature (WQ) rules are defined over the whole parameter space. Thus, they take advantage of the continuity of the entire support, which is violated by \bmchange{the interface $\TrimCurve$ present in the cut support $\supportdomainCut_{{\indexA}}$.}
Discontinuous weighted quadrature (DWQ) overcomes this problem \cite{Marussig2022}.


\bmchange{For outlining the essential idea of DWQ, let us consider a univariate basis function intersected once by an interface $\TrimCurve$, as illustrated in \cref{fig:DWQconstr}.
By introducing an artificial discontinuity $\discCoeff$ at the knot between $\supportdomainReg_{{\indexA}}$ and $\supportdomainCut_{\indexA}$, the envisaged split of the support is incorporated into the parameter space.
This $\discCoeff$ reduces the smoothness only for the computation of the corresponding weighted quadrature rule; therefore, we label it artificial.
The resulting DWQ rule considers $\discCoeff$ as a discontinuity, i.e., the quadrature points within $\supportdomainReg_{{\indexA}}$ are independent of those within $\supportdomainCut_{\indexA}$.
As pointed out in \cref{sec:layout}, a reduced continuity within the spline space increases the required number of weighted quadrature points, which has to be taken into account during the computation of the DWQ rule. In particular, the} initial point layout of the WQ rules $\quadRule^{(0)}(\cdot)$ and the identified $\discCoeff$ are the starting point for setting up univariate DWQ rules $\quadRuleDisc^{(0)}(\cdot)$.
\bmchange{Following} \cref{fig:DWQconstr}, 
the construction employs the \bmchange{subsequent} steps:
\begin{enumerate}[label=(\alph*)]
    \item Knot insertion at $\discCoeff$ so that the univariate basis becomes $C^{-1}$ continuous there, 
    and storing of the corresponding subdivision matrix $\mathbf{S} \in \R^{\bmchange{\ndofsRefine}\times \bmchange{\ndofs}}$ where $\bmchange{\ndofs}$ and $\bmchange{\ndofsRefine}$ is the number of \bmchange{functions} of the initial and resulting discontinuous basis, respectively. See \cref{sec:knotInsertion} for details on setting up $\mathbf{S}$.
    \item Defining the location of the quadrature points $x_k$ with $k=1,\dots,n_k$ by taking the ones of the WQ rules $\quadRule^{(0)}(\cdot)$ and adding further nested points \bmchange{in the elements adjacent to  $\discCoeff$} to comply with the minimum number required for the exactness condition of the discontinuous basis.
    \item Computation of the quadrature weights $\tilde{\quadWeight}_{k,\indexA} \in \R^{n_k\times\bmchange{\ndofsRefine}}$ at the points $x_k$ for the basis functions $\tilde{\Bspline}_\indexA$ with $\indexA=1,\dots,\bmchange{\ndofsRefine}$, of the refined basis using \cref{eq:weightedQuadRuleSystem}, and subsequent multiplication 
    \begin{align}
        \bmchange{
        \quadWeight_{k,j} = \sum_{i=0}^{\ndofsRefine} \tilde{\quadWeight}_{k,i}\; \subDMatrix_{ij}} && \text{for}\quad j=1,\ldots , \ndofs\quad\text{and}\quad k=1,\dots,n_k 
    \end{align}
    to obtain the weights $\quadWeight_{k,\indexB} \in \R^{n_k\times \bmchange{\ndofs}}$ of $\quadRuleDisc^{(0)}(\cdot)$ for the initial smooth univariate test functions $\Bspline_\indexB$ with $\indexB=1,\dots,\bmchange{\ndofs}$. 
    \item Finally, the DWQ quadrature points within the \emph{cut} part, i.e.,  $x_k\in\supportdomainCut_{\boldsymbol{\indexA}}$, can be neglected and replaced by an element-wise rule.
\end{enumerate}
\bmchange{This procedure provides the $\quadRuleDisc^{(0)}(\cdot)$ rules for all cut basis functions that share the same $\discCoeff$ value. The extension to artificial artificial  discontinuities within a single univariate support follows the same logic.}
Note that the number of quadrature points $n_k$ is higher for $\quadRuleDisc^{(0)}(\cdot)$ than for $\quadRule^{(0)}(\cdot)$.  
These additional points account for the discontinuity due to $\discCoeff$ and enable the substitution of the weighted quadrature points with an element-wise one in the cut region.

\bmchange{
\subsubsection{Computation of discontinuous weighted quadrature rules $\quadRuleDisc^{(1)}(\cdot)$}
}
As demonstrated in \cite{Hiemstra2019}, \bmchange{the} weighted quadrature \bmchange{rule $\quadRule^{(1)}_{i,p}(\cdot)$} for the first derivative of \bmchange{the $i$th} test function \bmchange{of degree $p$} can be computed based on \cref{eq:Bspline_Np_Deriv} leading to the following linear combination 
\begin{equation}
	\label{eq:WQ_Deriv}
	\quadRule^{(1)}_{i,p}(v)  = \frac{p}{\uu_{i+p}-\uu_{i}} \: 
    \quadRule^{(0)}_{i,p-1}(v) 
				    - \frac{p}{\uu_{i+p+1}-\uu_{i+1}} \: \quadRule^{(0)}_{i+1,p-1}(v).
\end{equation}
\bmchange{where $\quadRule^{(0)}_{i,p-1}(\dots)$ are the weighted quadrature rules for the test functions of lower degree $p-1$.}
Here, we adopt this strategy to compute discontinuous weighted quadrature rules $\quadRuleDisc^{(1)}(\cdot)$.
For the following computation the subdivision matrix $\mathbf{S}$ of \cref{sec:DWQcomputation} can be reused since knot insertion at the identified artificial \bmchange{discontinuity} $\discCoeff$ is again the starting point. 
\bmchange{Without a loss of generality, we again detail the construction steps for introducing a single $\discCoeff$ into the support} 
\begin{enumerate}[label=(\alph*)]
    \item Splitting the basis at \bmchange{$\discCoeff$}
    into separate knot vectors $\tilde{\KV}_b$ with \bmchange{$b=1,2$.}
    \item Applying \cref{eq:WQ_Deriv} to each sub-basis \bmchange{induced by} $\tilde{\KV}_b$ to compute the related $\quadRule^{(1)}_{i,p,b}(\cdot)$ and their weights $\tilde{\quadWeight}^{(1)}_{k,\indexB,b}$.
    \item Multiplication of the block diagonal matrix obtained by \bmchange{$\diag\left(\tilde{\quadWeight}^{(1)}_{k,\indexB,1},\tilde{\quadWeight}^{(1)}_{k,\indexB,2}\right)$} 
    with $\mathbf{S}$ to obtain the weights ${\quadWeight}^{(1)}_{k,\indexA}$ of the entire $\quadRuleDisc^{(1)}(\cdot)$.
\end{enumerate}
Finally, it is noted that the quadrature points of $\quadRuleDisc^{(1)}(\cdot)$ that are within the cut part, i.e.,  $x_k\in\bmchange{\supportdomainCut_{\indexA}}$, can be again replaced by element-wise quadrature rules without affecting the integration result in \bmchange{$\supportdomainReg_{{\indexA}}$}. 

\bmchange{
\subsection{Detection of artificial discontinuities for multivariate splines}
\label{sec:detectdisc}
}

\bmchange{Although sum factorization relies on univariate integrals, 
constructing DWQ rules for multivariate splines requires further considerations. 
To be precise, finding proper positions of the artificial discontinuities becomes more involved.}
In the multivariate case, an artificial discontinuity $\discCoeff_{\iSubDim}$ \bmchange{introduced for the parametric direction $\iSubDim$ of an $\pdim$-dimensional background mesh} propagates through the entire parameter space due to the tensor product structure of the spline basis\bmchange{. Thus, each $\discCoeff_{\iSubDim}$ represents a line or a plane cutting through the 2D or 3D domain, respectively, which
divides a domain into sub-domains that also possess a tensor product structure.
Since an interface $\TrimCurve$ does not generally follow this structure, it is usually not possible to completely separate $\supportdomainReg_{\boldsymbol{\indexA}}$ and $\supportdomainCut_{\boldsymbol{\indexA}}$ by these $\discCoeff_{\iSubDim}$.  
Instead, the regular part of a support is further split, i.e., 
$\supportdomainReg_{\boldsymbol{\indexA}}=\supportdomainDWQ_{\boldsymbol{\indexA}}\cup\supportdomainGQ_{\boldsymbol{\indexA}}$, where $\supportdomainDWQ_{\boldsymbol{\indexA}}$ denotes a larger tensor product subdomain that can be subject to (discontinuous) weighed quadrature, and $\supportdomainGQ_{\boldsymbol{\indexA}}$ represents the remaining non-cut interior elements of the support.
Note that $\supportdomainGQ_{\boldsymbol{\indexA}}$ does not occur in the univariate case. These regions are treated element-wise and integrated by sum factorization with standard Gauss quadrature. Hence, we label the related elements as \emph{Gauss elements}.

To sum up,
}
the placement of $\discCoeff_{\iSubDim}$ affects what kind of quadrature (i.e., Gauss, WQ, or DWQ) is required in which element of the background mesh and is therefore essential for the overall number of quadrature points required. 
For the sake of clarity, the explanations and examples will consider the 2D case, but all concepts extend straightforwardly to the 3D setting.
First, we discuss the strategy proposed in \cite{Marussig2022a} that determines the most effective choice of $\discCoeff_{\iSubDim}$ for each cut test function independently.
Then a novel approach is presented that specifies all $\discCoeff_{\iSubDim}$ for all cut test functions at once.

\subsubsection{Individual placement for each test function}
\label{sec:discknotlocal}

\begin{figure}[t]
    \centering
    \includegraphics[scale=1.0]{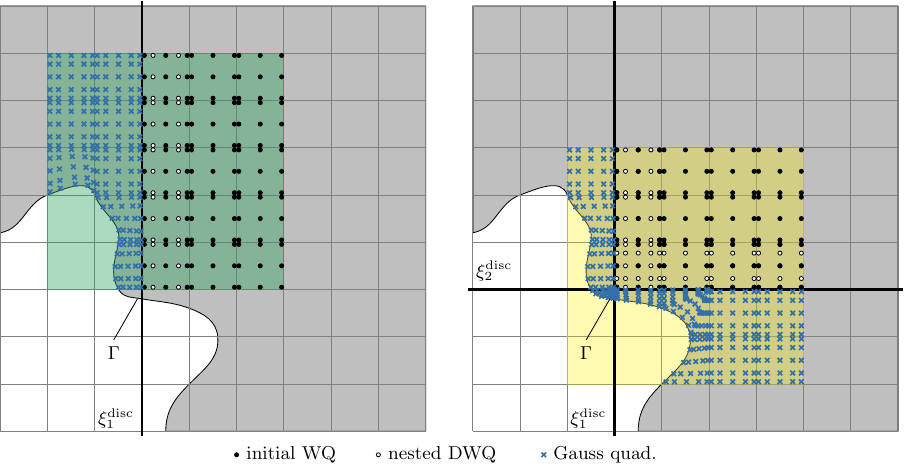}
    \caption{Individual detection of artificial discontinuities $\discCoeff_{\iSubDim}$ for cut test functions. The artificial discontinuities $\discCoeff_{\iSubDim}$ are chosen for each function separately to minimize Gauss points (blue crosses). Due to $\discCoeff_{\iSubDim}$, the initial WQ points (black dots) require additional ones (white dots).}
    \label{fig:DWQlocal}
\end{figure}

This concept aims to find the knot values $\discCoeff_{\iSubDim}$ \bmchange{yielding the largest $\supportdomainDWQ_{\boldsymbol{\indexA}}$} of each cut B-spline $\Bspline_{\boldsymbol{\indexA}}$.
Thereby, a maximum of one $\discCoeff_{\iSubDim}$ per direction\bmchange{, $\iSubDim=1,\dots,\pdim$,} is introduced within the support $\Supp{ \Bspline_{{\boldsymbol{\indexA}}} }$. 
\bmchange{Note that it would be possible to define several $\discCoeff_{\iSubDim}$ per direction, but every additional $\discCoeff_{\iSubDim}$ results in more nested DWQ points, and thus, such a discontinuous weighted quadrature rule becomes more expensive.}
\Cref{fig:DWQlocal} sketches the situation for two test functions, where the individual $\discCoeff_{\iSubDim}$ divide the $\Supp{ \Bspline_{{\boldsymbol{\indexA}}} }$ into two and 4 parts, respectively. 
The parts \bmchange{that} contain cut elements employ Gaussian quadrature \bmchange{since they represent the union of $\supportdomainCut_{\boldsymbol{\indexA}}$ and $\supportdomainGQ_{\boldsymbol{\indexA}}$. The others define $\supportdomainDWQ_{\boldsymbol{\indexA}}$ and} can be integrated by DWQ rules.
Note the nested DWQ points added \bmchange{within $\supportdomainDWQ_{\boldsymbol{\indexA}}$} next to the parametric lines associated with $\discCoeff_{\iSubDim}$.
When applied to 3D domains, 
each $\discCoeff_{\iSubDim}$ introduces a plane that splits the cut basis function's support. When only a single $\discCoeff_{\iSubDim}$ is used per test function, the extension to 3D is straightforward \bmchange{because determining the best location only requires counting interior elements on one side of $\discCoeff_{\iSubDim}$.}
Considering multiple $\discCoeff_{\iSubDim}$, however, \bmchange{makes the implementation more involved since different subregion types occur, i.e., rectangles in 2D and cubes in 3D.}

This strategy is optimal \bmchange{regarding the} number of quadrature points per cut test function.
At the same time, it may be sub-optimal for the overall integration of the background mesh because the overlap of the individual $\discCoeff_{\iSubDim}$ can result in elements that require \bmchange{weighted} and Gauss quadrature schemes.
When choosing the weighted quadrature points as \bmchange{a} subset of the Gaussian ones, as discussed in \cref{sec:layout}, the problem is mitigated since the number of points is bounded by the Gauss rule.
Nevertheless, the affected elements have a computational overhead \bmchange{in the numerical integration since they are evaluated by element assembly \emph{and} row assembly.} 

\bmchange{\subsubsection{Placement based on a global interface}}
\label{sec:discGlobal}

Alternatively to the previous strategy, one can consider the \bmchange{whole} parameter space \bmchange{at once} rather than its \bmchange{individual} cut test functions. 
Here, we propose \bmchange{introducing a global interface within the valid domain that guides the selection of artificial discontinuities. The goal is to have a general split between Gauss elements and those evaluated by weighted quadrature so that no region is integrated by two different routines.}

The starting point is a global partitioning of the background mesh into boxes of width $h_b$, where $h_b$ defines how many elements per direction fit into a box.
\bmchange{The boundaries of these boxes specify the possible locations of artificial  discontinuities $\discCoeff_{\iSubDim}$. Thus, this}
box partition must be aligned with the elements and \bmchange{enclose} the background mesh. 
\bmchange{The first step to selecting suitable $\discCoeff_{\iSubDim}$ is the identification of admissible and inadmissible regions.}
In particular, \emph{admissible} boxes possess only interior elements, while \emph{inadmissible} ones either contain the  boundary $\TrimCurve$ or are entirely outside the domain $\patchdomain$.
The interface $\Gamma^{\textnormal{disc}}$ between these different types of blocks determines the regions integrated by weighted quadrature rules $\patchdomain_{WQ}$ and Gauss rules $\patchdomain_{GQ}$.  
To be precise, $\patchdomain_{WQ}$ represents the elements within admissible blocks, while $\patchdomain_{GQ}$ contains all remaining elements of the valid computational domain $\patchdomain$.
Hence, $\patchdomain_{GQ}$ is the collection of all cut and Gauss elements.
Using $\Gamma^{\textnormal{disc}}$ and the overlap $\bmchange{\supportdomainDWQglobal_{\boldsymbol{\indexA}}} \coloneqq  \Supp{ \Bspline_{{\boldsymbol{\indexA}}} }  \cap \Blue {\patchdomain_{WQ}} \Black$, we can adapt the \bmchange{classification} of a test function $\Bspline_{{\boldsymbol{\indexA}}}$ to:
\begin{itemize}
        \item \emph{Exterior} if $\supportdomain_{\boldsymbol{\indexA}} = \emptyset$, 
        \item \emph{Interior} if $\bmchange{\supportdomainDWQglobal_{\boldsymbol{\indexA}}} = \Supp{ \Bspline_{{\boldsymbol{\indexA}}} }$, 
        \item \emph{Cut} if  $0 < \left|\bmchange{\supportdomainDWQglobal_{\boldsymbol{\indexA}}}\right| <\left| \Supp{ \Bspline_{{\boldsymbol{\indexA}}} }\right|$,
        \item \bmchange{\emph{Gauss} if $\supportdomainDWQglobal_{\boldsymbol{\indexA}}=0$ but $\supportdomain_{\boldsymbol{\indexA}} > \emptyset$.}
\end{itemize}
Note that the definition of exterior test function stays the same as presented in \cref{sec:functionTypes}, while the others now consider $\Gamma^{\textnormal{disc}}$ rather than $\Gamma$.
\bmchange{
Furthermore, the Gauss class represents test functions that are only non-zero within $\patchdomain_{GQ}$ and, therefore, not subjected to weighed quadrature rules.
Hence, the subdomain $\supportdomainDWQglobal_{\boldsymbol{\indexA}}$ is like $\supportdomainDWQ_{\boldsymbol{\indexA}}$ in the individual placement approach but specified based on the global entity $\Gamma^{\textnormal{disc}}$.
}

\begin{figure}[t]
    \centering
    \includegraphics[scale=1.0]{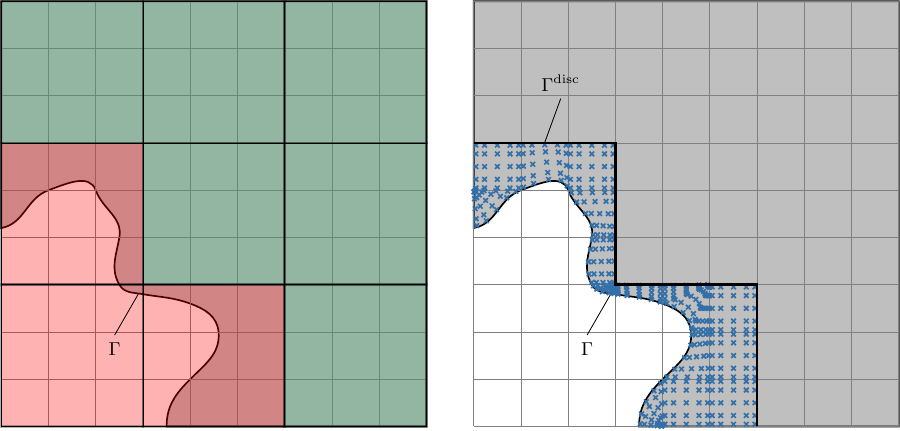}
    \caption{A box partition of a 2D background mesh using boxes of size $h_b=3$. The left figure shows the admissible blocks (green) and inadmissible blocks (red). The right figure depicts the corresponding interface $\Gamma^{\textnormal{disc}}$ and the Gauss quadrature points (blue crosses) within the inadmissible region.}
    \label{fig:DWQglobal}
\end{figure}
\Cref{fig:DWQglobal} shows an example of a 2D background mesh decomposed by boxes with $h_b=3$.
We highlight that inadmissible boxes may contain Gauss elements as well. 
It is beneficial to \bmchange{reduce} their number because integrating them with a weighted quadrature would be more efficient. 
\bmchange{Note that the degree $\pu$ does not play a role in this context, but}
one can alter $h_b$ or the origin of the box partition.
Considering the latter for a fixed $h_b$, all boxes can be shifted $h_b-1$ times in each direction -- one element at a time.
This procedure results in $(h_b-1)^{\bmchange{\pdim}}$ different partitions; the one with the smallest number of Gauss elements is preferred.

Furthermore, the parameter $h_b$ controls how well $\Gamma^{\textnormal{disc}}$ approximates $\TrimCurve$, which leads to two implications: 
(i) the smaller $h_b$, the smaller the number of Gauss elements within inadmissible boxes, 
and (ii) the \bmchange{larger} $h_b$, the smaller the number of artificial discontinuities $\discCoeff_{\iSubDim}$ introduced.
While we strive for the former, the latter is advantageous too, because fewer nested points are required for the DWQ rules. 
\bmchange{Suppose $h_b<3$, all elements within an admissible block are adjacent to the block's boundary. If such a block is part of a cut test function's support, all elements would require nested DWQ points, resulting in the same number of points as for Gauss quadrature.}
Hence, setting $h_b=3$ seems to be an appropriate choice to balance these two contradictory goals of minimizing the number of Gauss elements in inadmissible blocks and reducing the number of nested DWQ points.

\begin{figure}[t]
    \centering
    \includegraphics[scale=1.0]{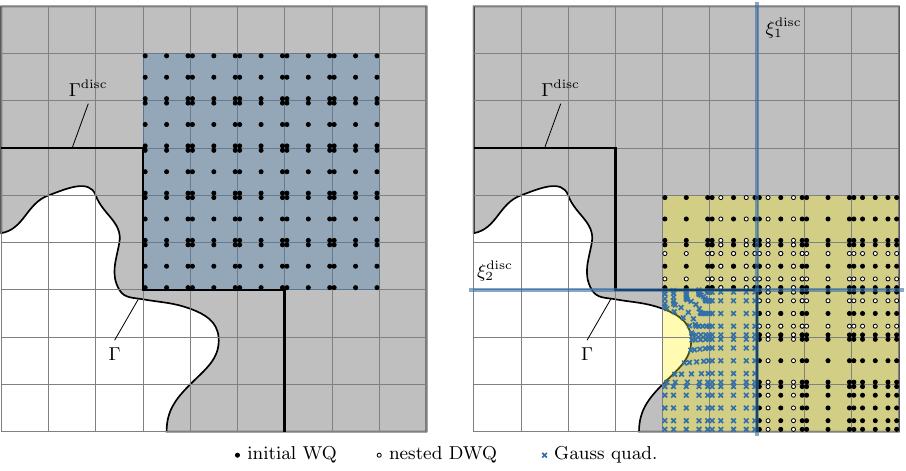}
    \caption{\bmchange{Weighted} quadrature point assignement based on the box partition interface $\Gamma^{\textnormal{disc}}$. Left: the test function's support (blue) does not intersect with $\Gamma^{\textnormal{disc}}$ allowing the use of WQ points. Right: the test function's support (yellow) intersects with $\Gamma^{\textnormal{disc}}$ yielding to a DWQ rule where $\discCoeff_{\iSubDim}$ are the extensions of $\Gamma^{\textnormal{disc}}$ within the support. \bmchange{The intersection and its extensions are indicated by blue lines.}}
    \label{fig:DWQglobalExamples}
\end{figure}

The final outcome of the partitioning approach is the interface $\Gamma^{\textnormal{disc}}$.
Test functions whose support intersects with $\Gamma^{\textnormal{disc}}$ are subjected to DWQ, and the intersections define its $\discCoeff_{\iSubDim}$ as sketched in \cref{fig:DWQglobalExamples}.
\bmchange{The left example shows an interior test function; here, the weighed quadrature points are given by the tensor product of the univariate layout, c.f.~\cref{fig:DWQlayout}(c).
The right example, on the other hand, represents a cut test function.}
Note that the edge of $\Gamma^{\textnormal{disc}}$ within the support represents the $\discCoeff_{\iSubDim}$ required for that test function, as indicated by \bmchange{the blue lines and the added nested DWQ points adjacent to them.}

Compared to the individual placement of $\discCoeff_{\iSubDim}$ described in \cref{sec:discknotlocal}, the number of cut test functions may increase since the approximation by $\Gamma^{\textnormal{disc}}$ can introduce superfluous Gauss elements, as discussed in the previous paragraphs. 
Indeed, \cref{fig:DWQglobalExamples} shows this circumstance since the lowest vertical edge of $\Gamma^{\textnormal{disc}}$ is adjacent to Gauss elements. 
Using boxes with variable $h_b$ may be an option, but the implementation gets more involved.
Furthermore, equally sized \bmchange{admissible} boxes have the same maximal number of quadrature points
\begin{align}
    n^b_q = \prod_{\indexC=1}^{\bmchange{\pdim}} (\pu_\indexC+1)\cdot2+\prod_{\indexC=1}^{\bmchange{\pdim}} n^i_q\cdot(h_b-2)
\end{align}
where $n^i_q$ is the number of \bmchange{univariate} weighted quadrature points \bmchange{per direction} within inner elements, \bmchange{e.g.}, $n^i_q=3$ for setting up a stiffness matrix.
\bmchange{Knowing the number of points in advance} is desirable since the metric and material-dependent part needs to be precomputed at the quadrature points for sum factorization. 
As highlighted in \cite{Hiemstra2019}, this is a drawback since the overall number of quadrature points $n_q$ has no upper bound with patch refinement. 
Therefore, the boxes can be used directly for balancing the computational and memory load, which is also beneficial when considering parallelization of the computation.

\begin{figure}[t]
    \centering
    \includegraphics[scale=1.0]{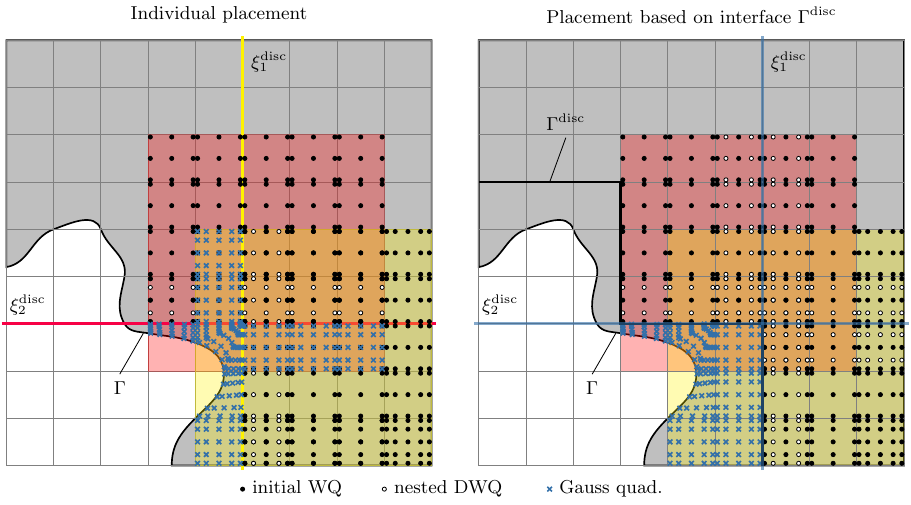}
    \caption{\bmchange{Quadrature points of partially overlapping cut test functions. Left: quadrature points due to an individual placement of $\discCoeff_d$. The color of the related lines indicates the test function $\discCoeff_d$ originates from. Right: Alternative point distribution for the same cut test functions due to the global concept.}}
    \label{fig:DWQglobalvslocal}
\end{figure}

\bmchange{We close this section by comparing the quadrature point layout of both discussed approaches for detecting artificial  discontinuities.
\Cref{fig:DWQglobalvslocal} depicts the quadrature points for two partially overlapping cut test functions. The individual placement concept yields an overlap of Gauss and weighed quadrature points. 
The number of affected elements correlates to the size of the cut test functions support and, thus, to the degree of the background mesh. 
Using the global interface $\Gamma^{\textnormal{disc}}$, on the other hand, prevents this behavior at the cost of introducing more nested discontinuous weighted quadrature points. Note that both basis functions share the same $\discCoeff_d$ since they intersect the same edges of $\Gamma^{\textnormal{disc}}$.}

\subsection{Estimation of the computational cost}

We recapitulate the number of floating point operations for setting up a 3D mass matrix of a tensor product spline of degree $\pu=\pu_1=\dots=\pu_{\bmchange{\pdim}}$ with maximal smoothness \cite{Hiemstra2019}: 
\begin{enumerate}[label=(\alph*)]
    \item $c\cdot\pu^9$ for element assembly and formation by looping over Gauss points,
    \item $c_1\cdot p^5 + c_2\cdot p^6 + c_3\cdot p^7$ for element assembly and sum factorization using Gauss quadrature,
    \item $c_1\cdot p^7 + c_2\cdot p^6 + c_3\cdot p^5$ for row assembly and sum factorization using Gauss quadrature, and
    \item $c_1\cdot p^4 + c_2\cdot p^4 + c_3\cdot p^4 $ for row assembly and sum factorization using weighted quadrature.
\end{enumerate}
For sum factorization, the different costs refer to the separated stages of the process -- one per parametric direction.
These estimates refer to the full, non-cut, spline discretization, i.e., the background mesh. 
However, these cost estimates can be associated with the immersed boundary method when using the combined placement of artificial discontinuities discussed in \cref{sec:discGlobal}.
In particular, (a) is assigned to cut elements, (b) refers to the costs related to the Gauss elements, and (d) applies to test functions subjected to WQ.
It remains to estimate the costs of DWQ for cut test functions.
Due to the nested DWQ points next to the artificial discontinuities $\discCoeff_{\iSubDim}$, there is a mix of $\pu+1$ and $n^i_q$ points within the related region
$\bmchange{\supportdomainDWQglobal_{\boldsymbol{\indexA}}}$.
If there is only a single $\discCoeff_{\iSubDim}$, the quadrature \bmchange{point} distribution is the same as for non-cut boundary test functions. 
In the case of multiple $\discCoeff_{\iSubDim}$, however, the worst-case scenario where all elements contain $\left(\pu+1\right)^{\bmchange{\pdim}}$ points is possible.
Thus, we consider (c) a conservative estimate for the cost of DWQ.
The number of floating point operations will often be less since
\bmchange{$\supportdomainDWQglobal_{\boldsymbol{\indexA}}$ covers only a subregion of the test function's support.}

To sum up, the cost for the fast formation and assembly for an immersed boundary method can be estimated by 
\begin{align}
    \begin{aligned}
        \label{eq:DWQcosts}
    &N_{WQ}\left(c_1\cdot p^4 + c_2\cdot p^4 + c_3\cdot p^4\right)\\ 
    +&
    N_{DWQ}\left(c_1\cdot p^7 + c_2\cdot p^6 + c_3\cdot p^5\right)\\
    +&N_{REG}\left(c_1\cdot p^5 + c_2\cdot p^6 + c_3\cdot p^7\right)\\
    +&N_{CUT}\left(c\cdot p^9\right)
    \end{aligned}
\end{align}
where $N_{WQ}$ and $N_{DWQ}$ are the number of interior and cut test functions, while $N_{REG}$ and $N_{CUT}$ refer to the number of Gauss and cut elements. 
Apparently, the cost of $N_{WQ}$ is the one with the lowest complexity, i.e., $\OfOrder\left(N_{WQ}\,\pu^{\bmchange{\pdim}+1}\right)$, which may be undermined by the other contributions with 
$\OfOrder\left((N_{REG}+N_{DWQ})\,\pu^{2\bmchange{\pdim}+1}\right)$ and $\OfOrder\left(N_{CUT}\,\pu^{3\bmchange{\pdim}}\right)$.
Luckily,
$N_{REG}$ and $N_{CUT}$ depend primarily on the boundary $\TrimCurve$\bmchange{, the interface $\Gamma^{\textnormal{disc}}$,} and the element size $h$ of the background mesh. Thus, they do not \bmchange{change} with $p$. \bmchange{$N_{DWQ}$ and $N_{WQ}$, on the other hand, increase and decrease under $p$-refinement, respectively, due to the enlargement of the test functions' supports. Hence, the number of test functions integrated by (discontinuous) weighted quadrature rules counteracts the contribution of $N_{CUT}$ to some extent. 
More importantly,}
$N_{CUT}$ does not scale with \bmchange{$h^{\pdim}$} but with the number of elements intersected by $\TrimCurve$, which has a lower dimension of approximately \bmchange{$h^{\pdim-1}$}.
Also, $N_{REG}$ and $N_{DWQ}$ increase slower with $h$ than $N_{WQ}$. 
Put differently, the contribution of $N_{WQ}$ becomes more dominant under $h$-refinement,
\bmchange{and the finer discretization, the more vital becomes efficiency.} 


\section{Elastostatics}

Let us shortly recap the problem considered, using the following notation.
We adopt the summation convention and use indices $i,j,k,l \in {1,2,3}$ to denote components in physical space. The partial derivative of a component $v_i$ of a vector field $\vect{v}$ shall be denoted by $v_{i,j}$. In addition, we consider the symmetric gradient, which shall be denoted by $v_{(i,j)} : = \tfrac{1}{2}(v_{i,j} + v_{j,i})$.

\subsection{Model problem}
Let $\bmchange{\patchdomain} \subset \Box$ denote the three-dimensional \emph{physical domain} \rhchange{and $\Box$ a bounding box referred to as the \emph{uncut domain}, which is the smallest enclosing rectangular domain. The physical domain $\bmchange{\patchdomain}$} is an open set with piecewise smooth boundary $\Gamma = \overline{\Gamma_{g_k} \cup \Gamma_{t_k}}$, with $\Gamma_{g_k} \cap \Gamma_{t_k} = \emptyset$ and outward unit normal vector $\vect{n}$. We consider displacements in a space $\vect{V}$ with components in $H^1(\bmchange{\patchdomain})$ and let $\vect{V}_g \subset \vect{V}$ denote the subspace that satisfies the Dirichlet conditions on $\Gamma_{g_k}$, $k=1,2,3$. In particular, $\vect{V}_0$, denotes the space with homogeneous Dirichlet conditions.

Let the displacement be decomposed as $\vect{V}_g \ni \vect{u} = \vect{v} +\vect{g}$, with $\vect{v} \in \vect{V}_0$ and $\vect{g} \in \vect{V}_g$. The weak form of the elastostatics problem seeks $\vect{v} \in \vect{V}_0$ such that
\begin{subequations}
\begin{align}
	a(\vect{v}, \vect{w}) = l(\vect{w}) - a(\vect{g}, \vect{w}) \quad \forall \vect{w} \in \vect{V}_0
\end{align}
with the bilinear and linear form given by
\begin{align}
	a(\vect{v}, \vect{w}) &= \int \nolimits_{\bmchange{\patchdomain}} w_{(i,j)} \, c_{ijkl} \; v_{(k,l)} dx & \vect{v}, \vect{w} \in \vect{V}\\
	l(\vect{w}) &= \int \nolimits_{\bmchange{\patchdomain}} w_{i} \, f_i \, dx  \; + \; \sum_{i=1}^3 \int \nolimits_{\Gamma_{t_i}} w_{i} \, t_i \, ds & \vect{w} \in \vect{V}.
\end{align}
Here $f_i$ denote the components of a conservative force vector and $t_i = \sigma_{ij} \, n_j$ denotes the traction on the Neumann boundary. 
Furthermore, $c_{ijkl}$ denotes the fourth order stiffness tensor. 
In this article, we consider an isotropic material, which implies that $c_{ijkl}(x) = \mu(x) \left(\delta_{ik} \delta_{jl} + \delta_{il} \delta_{jk}  \right) + \lambda(x) \delta_{ij} \delta_{kl}$, see \cite{Hughes2000b}. 
In our testcases, the Lam\'e parameters $\mu$ and $\lambda$ are positive constants, given in terms of the Young's modulus $E$ and Poisson's ratio $\nu$ as $\lambda = \nu E / ((1+\nu)(1-2\nu))$ and \bmchange{$\mu = E / (2(1+\nu))$}.
\end{subequations}

\subsection{The discretized problem}
In the Galerkin method, we consider discrete displacement fields in a finite-dimensional subspace $\vect{V}^h \subset \vect{V}$ and consider functions $\vect{V}^h_{g} \subset \vect{V}^h$ that approximately satisfy the Dirichlet condition on $\Gamma_{g_k}$, $k=1,2,3$. We partition the boundary data into two parts that are treated differently. We consider
\begin{align}
\begin{cases}
	\Gamma_{{g}_k} \cap \partial \Box & \text{Strong imposition} \\
	\Gamma_{{g}_k} \setminus \, \partial  \Box & \text{Weak imposition}
\end{cases}
\end{align}
Weak \bmchange{imposition} of Dirichlet data can be done simply via a penalty method, or using Nitsche. 
In the considered benchmarks such a term is not needed and so we focus on strong imposition of Dirichlet data and, clearly, weak imposition of natural boundary conditions, which is straightforward on fictitious domains.

We consider the discrete space with homogeneous boundary conditions, $\vect{V}^h_0$. For each component $k \in 1,2,3$, we consider an expansion in terms of basis functions 
\begin{align}
	v^h_k(x) &= \rhchange{\sum_{\idx{i}=1}^{N_k}} v^{(k)}_{\idx{i}} \, N^{(k)}_{\idx{i}}(x).
\end{align}
The basis functions, $N^{(k)}_{\idx{i}}(x)$, are linear combinations of B-splines, determined via the extended B-spline concept proposed in \cite{Hoellig2002,Hoellig2003}. It is shown therein that the resulting set of shape functions are linearly independent and have good stability properties, with respect to small cut elements. Extended B-splines may be conveniently implemented via a spline extraction matrix $\vect{C}^{(k)} \in \mathbb{R}^{\rhchange{N_k \times M}}$
\begin{align}
	N^{(k)}_{\idx{i}}(x) = \sum_{\idx{j}=1}^{M}  \vect{C}^{(k)}_{\idx{i} \idx{j}} \, B_{\idx{j}}(x)
\end{align}
Here $N_k \leq M$ denotes the dimension of the space of the $k$th component of displacement and $M$ denotes the dimension of the spline space on the uncut domain $\Box$. The entries in $\vect{C}^{(k)}_{\idx{i} \idx{j}}$ are determined using the approach in \cite{Hoellig2002,Hoellig2003} such that all polynomials of degree $p$ are preserved. Homogeneous boundary conditions on $\Gamma_{{g}_k} \cap \partial \Box$ are implemented, simply by the condition that all polynomials need to be preserved which satisfy the homogeneous boundary condition.

With this in place, the Galerkin method seeks $\vect{v}^h \in \vect{V}^h_0$ such that
\begin{align}
	a(\vect{v}^h, \vect{w}^h) = l(\vect{w}^h) - a(\vect{g}^h, \vect{w}^h) \quad \forall \vect{w}^h \in \vect{V}^h_0.
\end{align}
The computed displacement is then determined as $\vect{u}^h = \vect{v}^h + \vect{g}^h$. 

\subsection{Application of sum factorization}

The application of sum factorization and weighted quadrature to this problem follows exactly the description provided in \cite{Hiemstra2019} and is, thus, not repeated here.
The only difference is that each cut test function may consist of a set of tensor product regions that possess different combinations of WQ and DWQ weights.
Hence, the preparation of extracting the correct non-zero weights requires more attention.
Once the interior and cut test functions are assembled, the element contribution of the cut and Gauss elements are added as well.

\begin{remark}
    When using the individual placement of the artificial discontinuities, it has to be stored what Gauss element is associated with which cut test function. 
    In the scheme \bmchange{with the global interface}, this is not needed.
\end{remark}

\sisetup{ round-mode = places, round-precision = 2} 
\section{Numerical results}

\begin{figure}[h]
    \centering
    \begin{subfigure}[c]{0.48\textwidth}
        \centering
        \includegraphics[height=5.4cm]{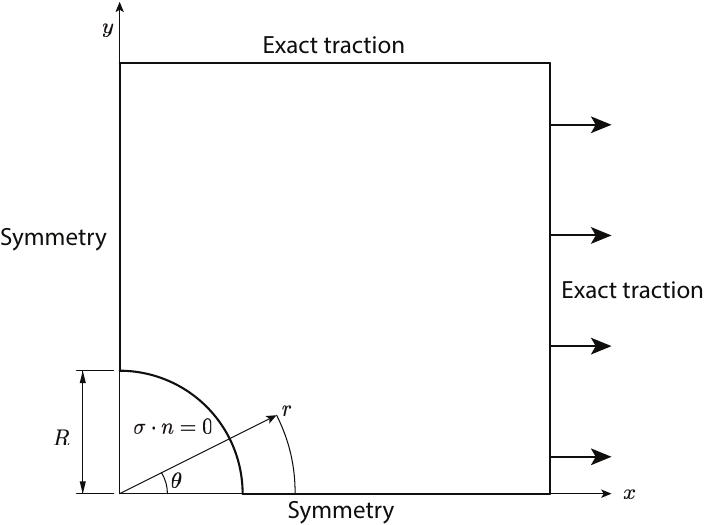}
        \subcaption{Hole in plate}
        \label{fig:benchmarkHole}
    \end{subfigure}
    \begin{subfigure}[c]{0.48\textwidth}
        \centering
        \includegraphics[height=5.4cm]{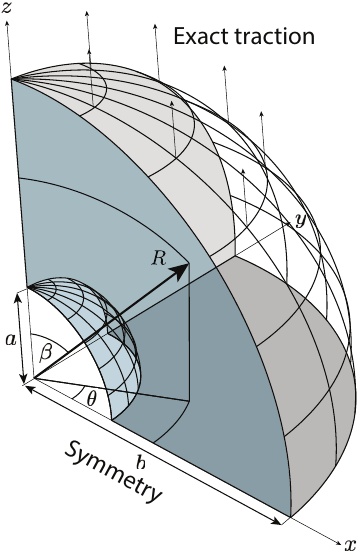}
        \subcaption{Spherical cavity}
        \label{fig:benchmarkCavity}
    \end{subfigure}
    \caption{The considered benchmark problems for linear elasticity}
    \label{fig:benchmarkAll}
\end{figure}

In this section, we investigate the efficiency of the proposed assembly and formation technique. 
Therefore, we revisit the numerical benchmark problems for linear elasticity shown in \cref{fig:benchmarkAll}, which have been used in \cite{Hiemstra2019} for boundary-fitted isogeometric discretization. Here, we apply them to the immersed boundary setting where the background mesh is defined by splines with maximal smoothness, and the cutting boundary $\TrimCurve$ is specified by a level set function $\phi$.

The implementation for the numerical experiments is built upon the feather-ecosystem\footnote{\href{https://gitlab.com/feather-ecosystem}{https://gitlab.com/feather-ecosystem}, \bmchange{ImmersedSplines v0.5.0,} accessed 8.6.2023}, which is a multiphysics isogeometric environment written in The Julia Programming Language \cite{bezanson2017julia}.
At its core, it relies on a sum factorization implementation optimized for element-by-element formation.
Hence, it provides a competitive reference solution concerning timings.
Besides, the code of the proposed discontinuous weighted quadrature utilizes the same representations as the feather-ecosystem \bmchange{and uses} its functions whenever possible to obtain a fair comparison.
Cut elements are integrated by a Julia wrapper to the Algoim\footnote{%
\bmchange{\href{https://github.com/JuliaBinaryWrappers/algoim\_jll.jl}{https://github.com/JuliaBinaryWrappers/algoim\_jll.jl}, v0.1.0, accessed 26.7.2023}
} library -- a C++ implementation of the routines presented in \cite{Saye2015,Saye2022}. 
\bmchange{Like} Gauss quadrature rules, the number of integration points increases with the degree by $\OfOrder\left(p^{\bmchange{\pdim}}\right)$; hence, the same cost estimates apply.

The timings are measured with the Julia package TimerOutputs.jl\footnote{\href{https://github.com/KristofferC/TimerOutputs.jl}{https://github.com/KristofferC/TimerOutputs.jl}, \bmchange{v0.5.23}, accessed 8.6.2023}, which has a small overhead in timing a code section (0.25$\mu s$ according to the package's documentation). 
The reported timings are, however, large enough not to be spoilt by this noise.
\bmchange{All computations are performed in serial, and the timings are the average of three successive runs.}
We measure the total timings for \bmchange{three different approaches:}
\begin{itemize}
    \item \emph{Element assembly}: 
    \bmchange{integration of all interior elements} by element-by-element formation using standard Gauss quadrature rules and sum factorization
    \item \emph{Row assembly (individual)}: integration of the regular region using the proposed fast formation with the individual placement of artificial discontinuities detailed in \cref{sec:discknotlocal}
    \item \emph{Row assembly (\bmchange{global})}: integration of the regular region using the proposed fast formation with the placement of artificial discontinuities \bmchange{based on a global interface, as detailed in \cref{sec:discGlobal},} employing boxes of size $h_b=3$
\end{itemize}
\bmchange{The first approach is an efficient element-wise technique and serves as a competitive reference implementation; the others are variances of the fast formation and assembly techniques. 
For the overall simulation process, these three approaches require an additional routine for treating of $\supportdomainCut_{\boldsymbol{\indexA}}$, which will be measured independently by}
\begin{itemize}
    \item \emph{Cut elements}: integration of cut elements using Algoim
\end{itemize}
\bmchange{Furthermore, sub-components of these four main contributions are investigated.}
The total timings also include assembling the element/test function contributions to the global stiffness matrix. 
They, however, do not include setting up the right-hand side of the system or solving the final system since 
these tasks are preformed by the identical routines for all cases studied.

We will first present the benchmark problems and their corresponding results in the following two subsections. 
The third subsection discusses all results together, allowing us to highlight specific similarities and contrasts between the examples.

\subsection{Hole in plate problem}

We first \bmchange{consider} an infinite plate with a circular hole under constant in-plane tension in \bmchange{the} $x$-direction, $T_x=10$, at infinity. 
The problem description is given in \cref{fig:benchmarkHole}.
The Poisson's ratio $\nu=0.3$ and the Young's modulus $E=10^5$ specify the linear elastic material.
This classic benchmark has a smooth 2D solution (see, e.g., \cite{Hiemstra2019}), but we solve it as a 3D problem allowing us to use the \bmchange{same} routines for both benchmarks, which improves their comparability.
Using symmetry conditions, only a quarter of the plate is \bmchange{modeled} by a background mesh of size $4\times4\times1/4$ partitioned into $(n_{el},n_{el},3)$ number of elements per direction with $n_{el} = \left\{5,10,20,40\right\}$.
These partitions \bmchange{also provide} the knot values for the spline bases of maximal smoothness of degree ($p$,$p$,$p$) with $p = \left\{2,4,6,8\right\}$.
\begin{figure}[b!]
    \centering
    \begin{subfigure}[b]{0.34\textwidth}
        \centering
        \includegraphics[height=3.3cm]{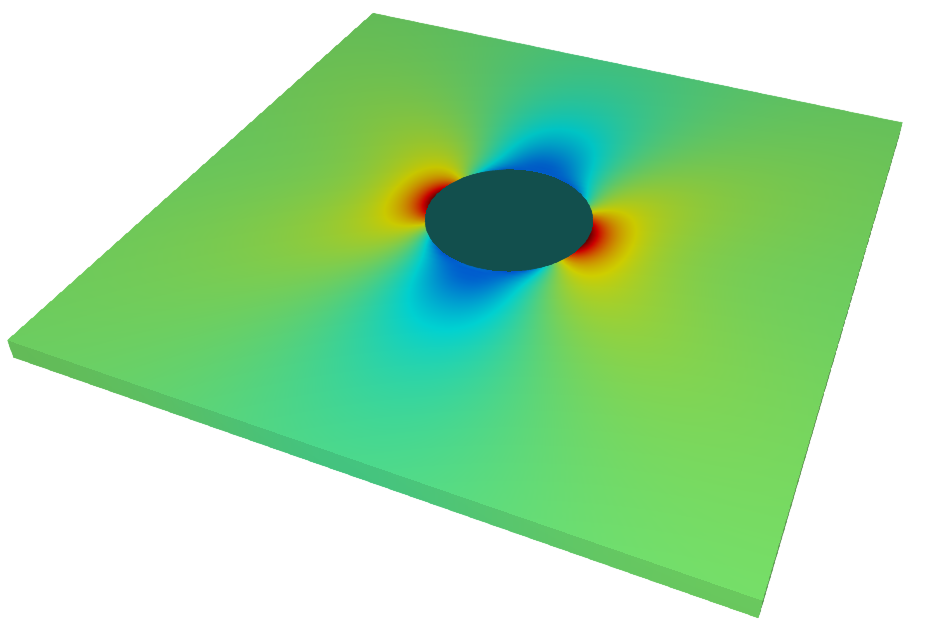}
        \subcaption{Stress component $\sigma_{xx}$ }
        \label{fig:HoleErrorPlot}
    \end{subfigure}
    \begin{subfigure}[b]{0.65\textwidth}
        \centering

        \begin{tabular}{ccc}
            \toprule            
            Component  & Element assembly & Row assembly (global) \\ \midrule
            $u_x$&\num{1.904789e-05} & \num{1.904789e-05} \\	
            $u_y$&\num{3.079523e-05} & \num{3.079523e-05} \\	
            $\sigma_{xx}$&\num{1.148239e-03} & \num{1.148239e-03} \\	
            $\sigma_{xy}$&\num{5.414056e-03} & \num{5.414056e-03} \\	
            $\sigma_{yy}$&\num{1.012643e-02} & \num{1.012643e-02} \\
            \bottomrule
        \end{tabular}
        \subcaption{Relative $L^2$ errors}
        \label{tab:HoleErrorTable}
    \end{subfigure}
    \caption{The numerical solution for the discretization of the hole in plate problem using $p=2$ and $n_{el}=40$: (a) distribution of the stress component $\sigma_{xx}$ shown for the entire plate and (b) the relative errors for the displacements $u$ and stresses $\sigma$ obtained by the reference element-by-element implementation and the proposed one using the \bmchange{global} placement scheme.}
    \label{fig:HoleError}
\end{figure}
The hole is introduced by the level set function $\phi=-x^2-y^2+1.0$, which cuts out a quarter circle of radius $R=1$ at the lower left corner of the background mesh.

\begin{figure}[t]
    \centering
    \includegraphics[scale=0.98]{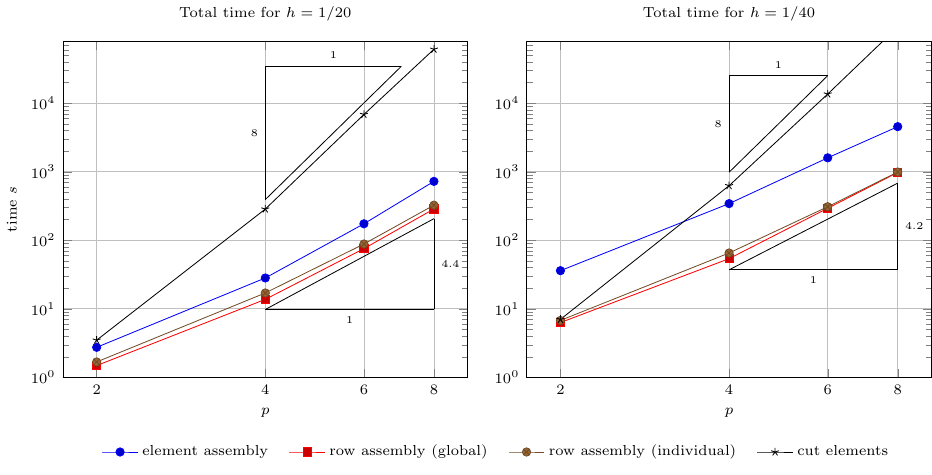}
    \caption{Hole in plate problem: absolute formation time in seconds for all degrees ($p,p,p$) considered and the two finest discretization, (left) $n_{el}=20$ and right $n_{el}=40$. }
    \label{fig:hpTimeTotal}
\end{figure}

\begin{figure}[H]
    \centering
    \includegraphics[scale=1.0]{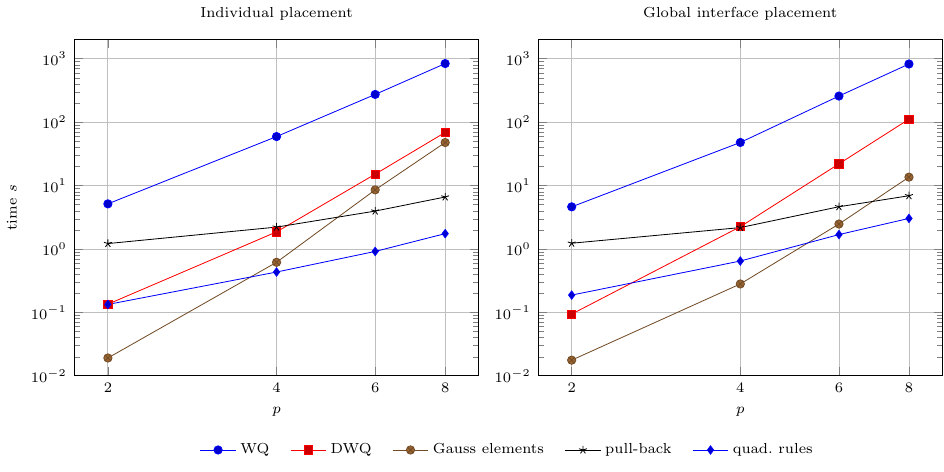}
    \caption{Hole in plate problem: various timing components of the row assembly strategies which only differ in the placement scheme (left/right) for the artificial continuities.}
    \label{fig:hpTimeRow}
\end{figure}

\begin{figure}[t]
    \centering
    \includegraphics[scale=1.0]{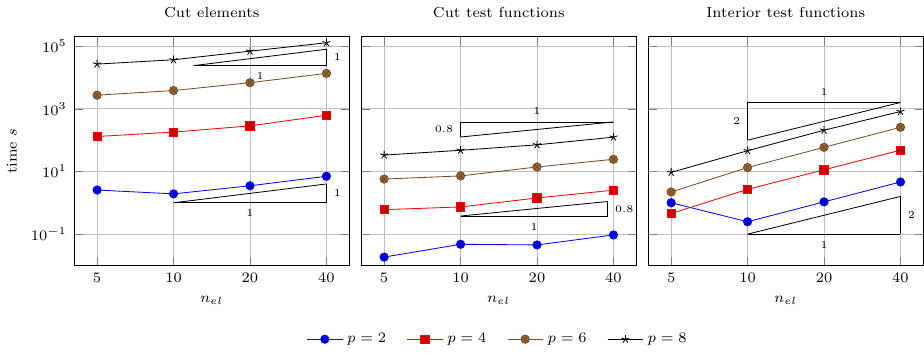}
    \caption{Hole in plate problem: different timings components of `row assembly (\bmchange{global})' and their development due to $h$-refinement of the background mesh which is expressed by the number of elements per direction $n_{el}$.}
    \label{fig:hpTimeH}
\end{figure}

An example of the obtained errors is summarized in \cref{fig:HoleError}.
Note that \cref{fig:HoleErrorPlot} displays the \bmchange{whole} plate, while only 1/4 was used for the simulation.
\Cref{fig:hpTimeTotal} illustrates the absolute formation time for the four key contributions.
The triangles indicate the complexity of the timings w.r.t.~the degree $p$.
Focusing on the finest discretization, a comparison of the different timing contributions of the row assembly methods is shown in \cref{fig:hpTimeRow}.
To be precise, the timings for applying WG, DWQ, and integrating the Gauss elements are reported. The remaining graphs refer to the pull-back of the material data \bmchange{at} all active integration points \bmchange{to the parameter space \cite{Hiemstra2019}} and the computation of the weighted quadrature rules.  
Here, we focus again on the $p$-related evolvement.
In \cref{fig:hpTimeH}, on the other hand, the different timing contributions of the `row assembly (\bmchange{global})' scheme are related to the number of elements per direction $n_{el}$. 
Here, the `cut test functions' graphs include the integration with DWQ rules and the one for the Gauss elements.

\subsection{Spherical cavity problem}

\begin{figure}[b!]
    \begin{subfigure}[b]{0.35\textwidth}
        \centering
        \includegraphics[height=5cm]{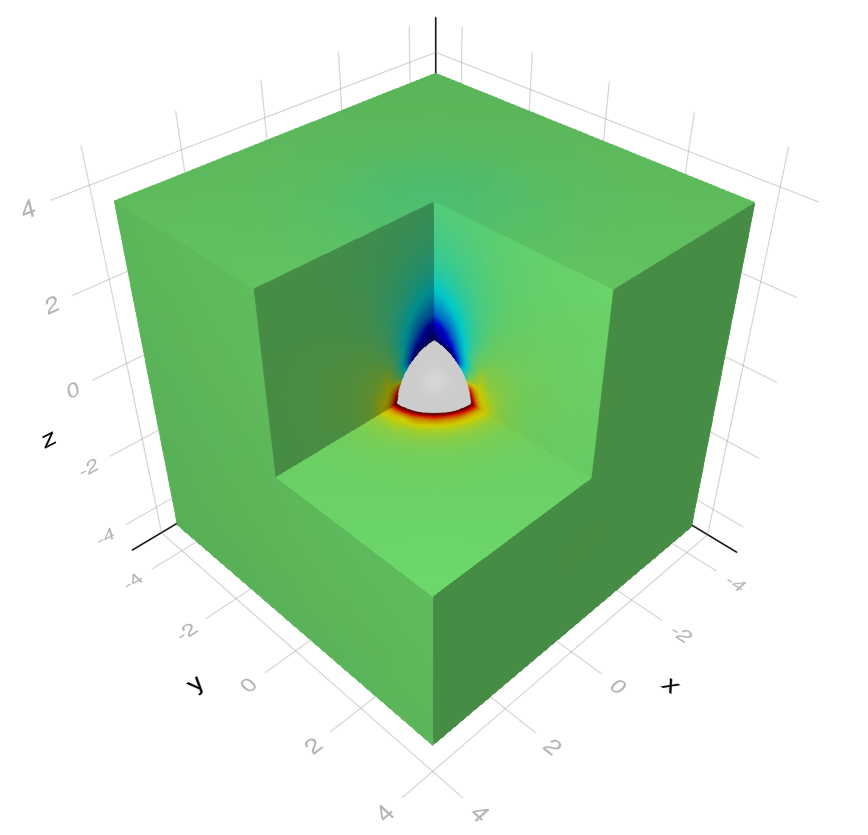}
        \subcaption{Stress component $\sigma_{zz}$ }
        \label{fig:CavityErrorPlot}
    \end{subfigure}
    \begin{subfigure}[b]{0.65\textwidth}
        \centering

        \begin{tabular}{ccc}
            \toprule            
            Component  & Element assembly & Row assembly (global) \\ \midrule        
        $u_{x}$&\num{1.605887e-02} & \num{1.605887e-02}\\	
        $u_{y}$&\num{1.607005e-02} & \num{1.607005e-02}\\	
        $u_{z}$&\num{1.058461e-02} & \num{1.058461e-02}\\	
        $\sigma_{xy}$&\num{7.682429e-01} & \num{7.682429e-01}\\	
        $\sigma_{xz}$&\num{7.004063e-01} & \num{7.004063e-01}\\	
        $\sigma_{yz}$&\num{7.001732e-01} & \num{7.001732e-01}\\	
        $\sigma_{zz}$&\num{4.541756e-02} & \num{4.541756e-02}\\
        \bottomrule
        \end{tabular}
        \subcaption{Relative $L^2$ errors}
        \label{tab:CavityErrorTable}
    \end{subfigure}
    \caption{The numerical solution for the discretization of the spherical cavity problem using $p=2$ and $n_{el}=10$: (a) distribution of the stress component $\sigma_{zz}$ shown for the entire \bmchange{domain}, and (b) the relative errors for the displacements $u$ and stresses $\sigma$ obtained by the reference implementation and the proposed one using the \bmchange{global} placement scheme.}
    \label{fig:CavityError}
\end{figure}

We now apply the proposed approach to a full 3D problem defined by a spherical cavity located at the origin of an infinite domain subjected to uniform uniaxial tension in \bmchange{the} $z$-direction, $T_z=10$, at infinity, as illustrated in \cref{fig:benchmarkCavity}.
The material parameters are set to the Poisson's ratio $\nu=0.3$ and the Young's modulus $E=10^5$. 
We again refer to \cite{Hiemstra2019} for the analytic reference solutions.
To be precise, \cref{fig:benchmarkCavity} shows the description for a boundary-fitted spline discretization. 
In this work, the outer boundaries will be defined \bmchange{by} the non-cut boundary of the background mesh, which has the size $4\times4\times4$. 
The cavity of radius $a=1$ is given by $\phi=-x^2-y^2-z^2+1.0$.
The spline discretizations \bmchange{investigated} have degree ($p$,$p$,$p$) with $p=\left\{2,4,6,8\right\}$ and 
number of elements per direction $(n_{el},n_{el},n_{el})$ with $n_{el}=\left\{5,10,21\right\}$. 
\bmchange{Case $n_{el}=20$ led to errors for degree 8 in subroutines related to the integration of cut element and has therefore been skipped.}

\begin{figure}[t]
    \centering
    \includegraphics[scale=1.0]{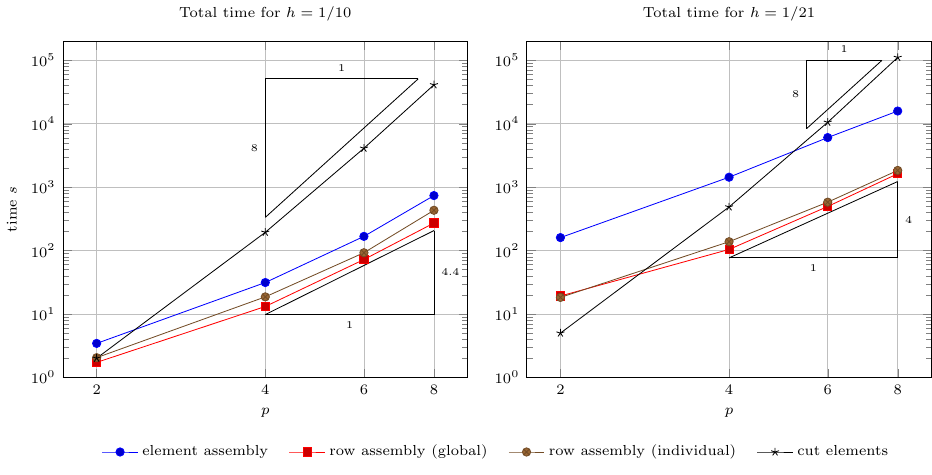}
    \caption{Spherical cavity problem: absolute formation time in seconds for all degrees ($p,p,p$) considered and the two finest discretization, (left) $n_{el}=10$ and right $n_{el}=21$.}
    \label{fig:CavityTimeTotal}
\end{figure}
\begin{figure}[H]
    \centering
    \includegraphics[scale=1.0]{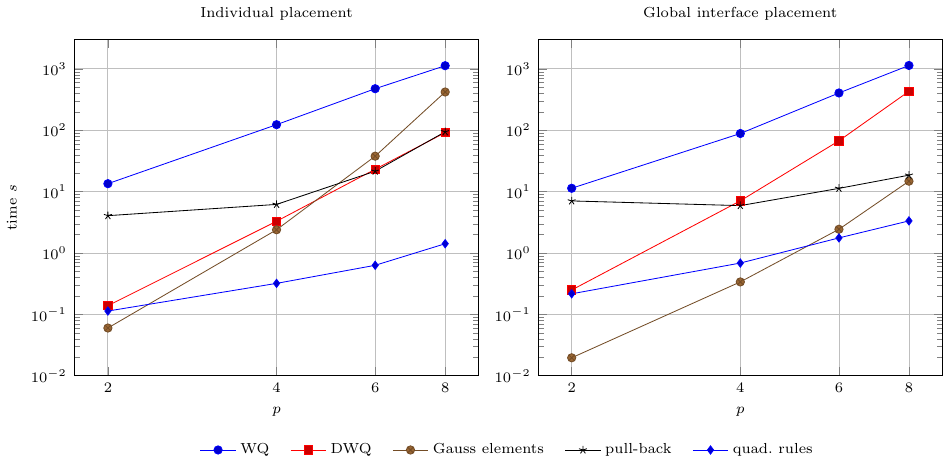}
    \caption{Spherical cavity problem: various timing components of the row assembly strategies which only differ in the placement scheme (left/right) for the artificial continuities.}
    \label{fig:CavityTimeRow}
\end{figure}

\Cref{fig:CavityError} reports the errors of the row assembly scheme and the reference implementation for an example discretization of the problem.
\Cref{fig:CavityErrorPlot} displays the entire domain, while the removed 1/8 of the cube is the actual computational domain.
The total formation times and the corresponding complexity are illustrated in \cref{fig:CavityTimeTotal},
and \cref{fig:CavityTimeRow} details the different contributions of the row assembly timings obtained for the finest discretization.

\begin{figure}[t]
    \centering
    \includegraphics[scale=1.0]{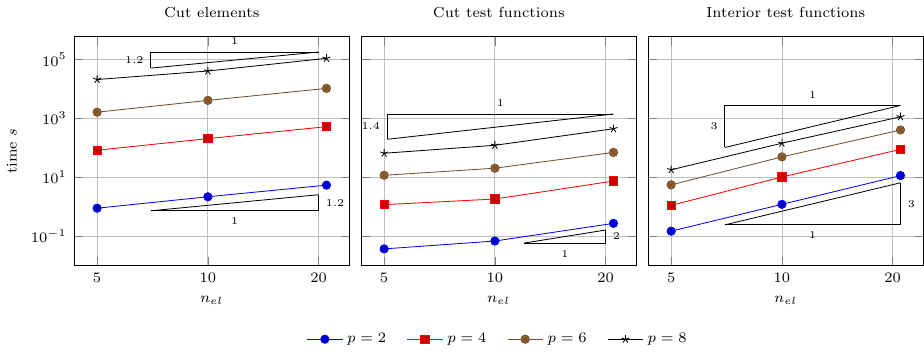}
    \caption{Spherical cavity problem: different timings components of `row assembly (\bmchange{global})' and their development due to $h$-refinement of the background mesh which is expressed by the number of elements per direction $n_{el}$.}
    \label{fig:CavityTimeH}
\end{figure}

\begin{figure}[h]
    \centering
    \includegraphics[scale=1.0]{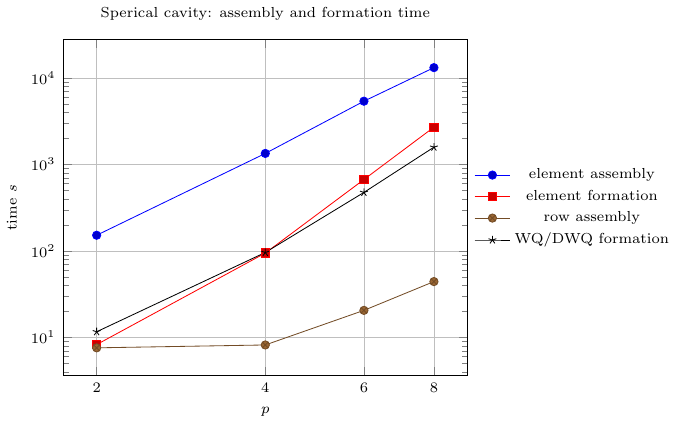}
    \caption{Spherical cavity problem: timings for the formation of stiffness contributions and their assembly to the global matrix: `element formation' utilized Gauss rules for element-by-element sum factorization, \bmchange{whereas} `WQ/DWQ formation' employs (\bmchange{discontinuous}) weighted quadrature rules.}
    \label{fig:CavityAssemblyVsFormation}
\end{figure}

Furthermore, \cref{fig:CavityTimeH} focuses on the subparts of the `row assembly (\bmchange{global})' and their behavior w.r.t.~the number of elements $n_{el}$.
Finally, we split the total time into the contributions due \bmchange{to} the formation of the element or test functions contribution and the assembly to the global stiffness matrix. \Cref{fig:CavityAssemblyVsFormation} compares the element-by-element reference routines divided into `element assembly' and `element formation', and the row-based one presented consisting of `WQ/DWQ formation' and `row assembly'.
\bmchange{Hereby, formation refers to the computation of the element or test function contributions, and assembly denotes the assignment of these contributions to the global stiffness matrix.}
Recall that both formation schemes employ sum factorization.

\subsection{Discussion}

Here, we interpret the timing results reported in the previous two sections.
First, it is apparent that the integration of cut elements is the bottleneck, as shown in \cref{fig:hpTimeTotal} and \cref{fig:CavityTimeTotal}, where all related graphs have a slope of 8.
This rate would also be the performance obtained by a conventional element-by-element routine that loops over Gauss points.
At the same time, the results confirm that this cost contribution decreases with the fineness of the discretization. 
This conclusion can be drawn by comparing the coarser discretizations with the finer ones by looking at the left and right illustrations as well as opposing \cref{fig:hpTimeTotal} and \cref{fig:CavityTimeTotal}. 
It is noteworthy that the slope of the row assembly results decreases with the greater $n_{el}$, which indicates that the factor contributed by the most efficient part, i.e., the WQ of interior test functions, becomes dominant.
\cref{fig:hpTimeH} and \cref{fig:CavityTimeH} provide further support for this statement, not only for the row assembly with cut and interior test functions but, more importantly, also for the comparison with the integration of cut elements.

Let us now focus on the different row assembly schemes using an (i) individual or (ii) \bmchange{global} positioning of the artificial discontinuities $\discCoeff$.
\bmchange{There is a small difference when comparing the total timings. However, the distinguishing behavior surfaces when the different contributions are examined.}
The affected timings are reported in \cref{fig:hpTimeRow} and \cref{fig:CavityTimeRow} by the graphs related to `DWQ' and `\bmchange{Gauss} elements' since their sum refers to the time needed to integrate all cut test functions.
Especially for the spherical cavity problems, it can be seen that `DWQ' scales better for the individual placement, which can be attributed to the fact that this scheme introduces \bmchange{fewer} $\discCoeff$.
This circumstance is also beneficial for the time `quad.~rules' associated with setting up the weighted quadrature rules.
However, the `\bmchange{Gauss} elements' \bmchange{graph} scales much worse.
\bmchange{This contribution is the major weakness of the individual positioning, and it is hard to assess in general since it depends on the degree and the shape of the boundary $\TrimCurve$.}
For the \bmchange{global} approach, on the other hand, `DWQ' and `\bmchange{Gauss} elements' have virtually the same slope, which complies with the estimate \cref{eq:DWQcosts}.  
Note that the `pull-back' time is also better, which can be partly explained by the fact that the boxes of the \bmchange{global} construction can be directly used for partitioning the quadrature points into tensor product regions, \bmchange{whereas,} for the individual concept, this has to be generated separately.

We close this discussion by taking a closer look at the performance of the reference solution `element assembly' and the proposed `row assembly' using the \bmchange{global} strategy.
In particular, \cref{fig:CavityAssemblyVsFormation} divides the total time into a formation and assembly part. 
Here, it can be seen that the element formation scales worse with $p$ \bmchange{than the weighed quadrature one. 
Moreover, the element assembly is a dominant factor for the computation time, while the row assembly is almost neglectable. At the same time,}
it should be noted that the assembly of element contribution can be implemented more efficiently, for example, by utilizing the matrix structure. 
The reported timings do not take advantage of such concepts.

\section{Conclusion}

This work extends the fast assembly and formation approach to isogeometric immersed boundary methods in the context of linear elasticity.
The concept relies on the combination of (i) sum factorization, (ii) weighted quadrature, and (iii) row assembly, which \bmchange{reduces} the number of operations from $\OfOrder\left(p^{3\bmchange{\pdim}}\right)$ to $\OfOrder\left(p^{\bmchange{\pdim}+1}\right)$ for $\bmchange{\pdim}$-dimensional tensor product splines of degree $p$. 
For immersed discretizations, however, techniques (i) and (ii) must be adapted because the boundary $\TrimCurve$ destroys the exploited spline properties. 
To be precise, the test functions cut by $\TrimCurve$ need special attention.

Following \cite{Marussig2022,Marussig2022a}, the use of sum factorization for cut test functions can be partly realized by dividing their support into regular and cut subregions.
The regular subregions follow a tensor product structure, but they require special weighted quadrature rules, which we refer to as discontinuous weighted quadrature.
In this paper, we derive 
these rules for setting up stiffness matrices based on concepts presented in \cite{Hiemstra2019}.
Furthermore, we propose a more effective way to construct them, which also allows an estimation of the overall computational cost.
\bmchange{The approach introduces 
an interface $\Gamma^{\textnormal{disc}}$ that divides the domain of interest into a parts integrated by Gauss and weighted quadrature, respectively. This $\Gamma^{\textnormal{disc}}$ provides the possible artificial discontinuities which determine the discontinuous weighted quadratures required.}

The overall cost of the fast immersed boundary method is determined by (i) the number of cut elements, $N_{CUT}$, (ii) the number of cut test functions, $N_{DWQ}$, and associated regular elements, $N_{REG}$, and (iii) the number of interior test functions, $N_{WQ}$. 
The corresponding number of operations varies from $\OfOrder\left(N_{CUT}\,\pu^{3\bmchange{\pdim}}\right)$ over $\OfOrder\left((N_{DWQ}+N_{REG})\,\pu^{2\bmchange{\pdim}+1}\right)$ to  $\OfOrder\left(N_{WQ}\,\pu^{\bmchange{\pdim}+1}\right)$.
For coarse discretizations, the cut element contribution dominates the overall cost, but the influence of $N_{WQ}$ increases with $h$-refinement. 
Our numerical experiments of two linear elasticity benchmarks confirm this behavior.

Nevertheless, improving the complexity of $N_{CUT}$ is an interesting future research direction. 
The integration of any advancement in this regard into the presented methodology is straightforward since the integration of cut elements is independent of the other components.
Regarding weighted quadrature, the number of points can be reduced by including boundary points in the layout; however, the correct assignment of the boundary point contributions needs special attention. 
\section*{Acknowledgements}

Benjamin Marussig was partially supported by the SFB TRR361/F90 CREATOR funded by the German Research Foundations DFG and the Austrian Science Fund FWF.

\appendix

\section{Knot insertion and subdivision matrix}
\label{sec:knotInsertion}

Knot insertion denotes the refinement of a B-spline object by adding knots $\uuRefined$ into its knot vector $\KV$.
This procedure results in a nested spline space, and the subdivision matrix $\bmchange{\tilde{\subDMatrix}} \: : \; \mathbb{R}^{\ndofs} \mapsto \mathbb{R}^{\bmchange{\ndofsRefine}}$ encodes the coefficients of the initial coarse representation and the refined one.
Considering quadrature weights, we obtain the relation
\begin{align}
	\label{eq:knotInsertion1}
	\tilde{\quadWeight}_i = \sum_{j=1}^{\ndofs} \bmchange{\tilde{\subDMatrix}}_{ij}  \; \quadWeight_j && \text{for } i=1,\ldots , \bmchange{\ndofsRefine}
\end{align}
where $\tilde{w}_i$ refers to quadrature weights computed for the refined basis functions.
If only one knot is inserted, i.e., $\KVRefined = \KV \cup \uuRefined$, where $\uuRefined \in [\uu_\indexSpan,\uu_{\indexSpan+1})$, the non-zero entries of $ \bmchange{\tilde{\subDMatrix}}$ are determined by
\begin{align}
	&
	\label{eq:knotInsertion}
	\begin{cases}
		\bmchange{\tilde{\subDMatrix}}(\indexC,\indexC-1) &=  1-\alpha_\indexC 	\\
		\bmchange{\tilde{\subDMatrix}}(\indexC,\indexC) &=  \alpha_\indexC
	\end{cases}
	&
	&
   \alpha_\indexC =	\left\{ \begin{array}{c l}
			1  &  \indexC \leqslant \indexSpan-\pu \\
			\frac{\uuRefined - \uu_\indexC}{\uu_{\indexC+\pu} - \uu_\indexC } & \indexSpan-\pu+1  \leqslant \indexC \leqslant \indexSpan\\
			0  &  \indexC \geqslant \indexSpan+1 \\
			\end{array} 
		\right.
		&
\end{align}
Multiple knots can be inserted by repeating this process, and the multiplication of the individual single-knot matrices yields the overall subdivision matrix.
For introducing \bmchange{an artificial discontinuity $\discCoeff$} for the discontinuous weighted quadrature rules\bmchange{, the} knot $\discCoeff$ has to be inserted so that the multiplicity after knot insertion $\tilde{\multi}\left(\discCoeff\right)=p+1$. \bmchange{The subdivision matrix used to map values of the refined basis to the initial one is given by $\subDMatrix=\tilde{\subDMatrix}^\intercal$.}

\bibliographystyle{myplainnat}

\bibliography{fastimmersed}

\end{document}